\documentclass[a4paper, onecolumn, 12pt, accepted=2025-08-20]{quantumarticle}
\pdfoutput=1

\usepackage[utf8]{inputenc}
\usepackage{lipsum}

\usepackage{amsmath, amssymb,amsfonts,amsthm,mathtools}
\usepackage{physics}  % all braket notation at once
\usepackage{xspace,graphicx,relsize,bm,xcolor}
\usepackage{soul} % provides  s p a c i n g o u t, underlining and some derivatives such as overstriking and highlighting: https://ctan.mc1.root.project-creative.net/macros/generic/soul/soul.pdf

\usepackage{parskip}  % no indentation but some space between paragraphs

\usepackage{dsfont}
\usepackage[T1]{fontenc}

\usepackage[
    backend=biber,
    style=alphabetic,
    sorting=anyt,
    minalphanames=3,
    maxalphanames=3,
    maxnames=99,
    backref=true
    ]{biblatex}
\DefineBibliographyStrings{english}{%
  backrefpage = {page},% originally "cited on page"
  backrefpages = {pages},% originally "cited on pages"
}

\usepackage{hyperref}
\usepackage[margin=1in]{geometry}
\usepackage[singlespacing]{setspace}
\definecolor{linkcol}{rgb}{0.0,0.55,0.7}
\definecolor{citecol}{rgb}{0.0, 0.6, 0.45}
\definecolor{urlcol}{rgb}{0.7, 0.0, 0.55}
\hypersetup{
	colorlinks,
	linkcolor={linkcol},
	citecolor={citecol},
	urlcolor={urlcol}
}

\usepackage{url}

\usepackage{subcaption}
\usepackage{mleftright}
\usepackage{multirow}

\usepackage{algorithm}% http://ctan.org/pkg/algorithms
\usepackage{algpseudocode}% http://ctan.org/pkg/algorithmicx

\usepackage{cleveref} % nicer theorem referencing
\allowdisplaybreaks

\def\01{\{0,1\}}
\newcommand{\mc}[1]{\mathcal{#1}}
\newcommand{\mb}[1]{\mathds{#1}}
\renewcommand{\mathbb}[1]{\mathds{#1}}

\newcommand{\defeq}{:=}%{\coloneqq}

\DeclareMathOperator*{\Ex}{\mathbf{E}}
\let\Pr\relax
\DeclareMathOperator*{\Pr}{\mathbf{Pr}}
\DeclareMathOperator*{\Var}{\mathbf{Var}}

\newcommand{\eps}{\epsilon}

\newcommand{\A}{\ensuremath{\mathcal{A}}}
\newcommand{\D}{\ensuremath{\mathcal{D}}}

\newcommand{\Dn}{\mathcal{D}_n}
\newcommand{\G}{\ensuremath{\mathcal{G}}}

\newcommand{\f}{\ensuremath{\mathfrak{f}}}

\newcommand{\U}{\ensuremath{\mathrm{U}}}

\newcommand{\Cl}{\ensuremath{\mathrm{Cl}}}

\newcommand{\muC}{{\mu_{C}}}
\newcommand{\muU}{{\mu_{U}}}
\newcommand{\muS}{{\mu_{S}}}

\renewcommand{\eval}{\mathtt{Eval}}
\newcommand{\poly}{\mathrm{poly}}

\newcommand{\stat}{\mathtt{Stat}}

\newcommand{\kfrac}{\mathrm{frac}}

\newcommand{\tv}{\mathrm{d}_\mathrm{TV}}

\newcommand{\lr}[1]{\left(#1\right)}
\newcommand{\lrq}[1]{\left[#1\right]}
\newcommand{\lrb}[1]{\left\{#1\right\}}

\newtheoremstyle{mydefinitionsty}% 〈name〉
{10pt}% 〈Space above〉
{10pt}% 〈Space below〉
{}% 〈Body font〉
{}% 〈Indent amount〉1
{}% 〈Theorem head font〉
{}% 〈Punctuation after theorem head〉
{.5em}% 〈Space after theorem head〉2
{\textbf{\thmname{#1}~\thmnumber{#2}:  }\thmnote{(#3)}}% 〈Theorem head spec (can be left empty, meaning ‘normal’)〉
\theoremstyle{mydefinitionsty}
\newtheorem{definition}{Definition}
\newtheorem{remark}{Remark}
\newtheorem{note}{Note}

\newtheoremstyle{myproblemsty}% 〈name〉
{10pt}% 〈Space above〉
{10pt}% 〈Space below〉
{}% 〈Body font〉
{}% 〈Indent amount〉1
{}% 〈Theorem head font〉
{}% 〈Punctuation after theorem head〉
{.5em}% 〈Space after theorem head〉2
{\textbf{\thmname{#1}~\thmnumber{#2}:  }\thmnote{(#3)}}% 〈Theorem head spec (can be left empty, meaning ‘normal’)〉
\theoremstyle{myproblemsty}
\newtheorem{problem}{Problem}

\newtheoremstyle{mythmsty}% 〈name〉
{10pt}% 〈Space above〉
{10pt}% 〈Space below〉
{\itshape}% 〈Body font〉
{}% 〈Indent amount〉1
{}% 〈Theorem head font〉
{}% 〈Punctuation after theorem head〉
{.5em}% 〈Space after theorem head〉2
{\textbf{\thmname{#1}~\thmnumber{#2}:  }\thmnote{(#3)}}% 〈Theorem head spec (can be left empty, meaning ‘normal’)〉
\theoremstyle{mythmsty}
\newtheorem{inftheorem}{Informal Theorem}
\newtheorem{theorem}{Theorem}
\newtheorem{lemma}[theorem]{Lemma}
\newtheorem{corollary}[theorem]{Corollary}

\newenvironment{proofof}[1][\unskip]{%
\par
\noindent
\textit{Proof of #1:}
\noindent}
{\hfill$\qedsymbol$}

\newcommand{\ryantext}[1]{\textcolor{teal}{#1}}

\renewcommand{\ryantext}[1]{#1}

\definecolor{alexcolor}{rgb}{0.0, 0.47, 0.75}   % {0.27, 0.51, 0.71}  
\definecolor{todocol}{rgb}{0.8,0.2,0.0}

\definecolor{questioncolor}{rgb}{0.36, 0.54, 0.66}

\definecolor{aurometalsaurus}{rgb}{0.43, 0.5, 0.5}

\definecolor{green-munsell}{rgb}{0.0, 0.66, 0.47}

\title{On the average-case complexity of learning output distributions of quantum circuits}

\author{Alexander Nietner}
\affiliation{Author list in pseudorandom order. All authors contributed equally.}
\affiliation{Dahlem Center for Complex Quantum Systems, Freie Universit\"at Berlin, Germany}
\orcid{0000-0002-6685-8400}
\email{a.nietner@fu-berlin.de}

\author{Marios Ioannou}
\affiliation{Author list in pseudorandom order. All authors contributed equally.}
\affiliation{Dahlem Center for Complex Quantum Systems, Freie Universit\"at Berlin, Germany}
\email{marios.ioannou@fu-berlin.de}

\author{Ryan Sweke}
\affiliation{Author list in pseudorandom order. All authors contributed equally.}
\affiliation{IBM Quantum, Almaden Research Center, San Jose, CA 95120, USA}

\author{Richard Kueng}
\affiliation{Author list in pseudorandom order. All authors contributed equally.}
\affiliation{Institute for Integrated Circuits, Department of Computer Science, Johannes Kepler University Linz, Austria}

\author{Jens Eisert}
\affiliation{Author list in pseudorandom order. All authors contributed equally.}
\affiliation{Dahlem Center for Complex Quantum Systems, Freie Universit\"at Berlin, Germany}

\author{Marcel Hinsche}
\affiliation{Author list in pseudorandom order. All authors contributed equally.}
\affiliation{Dahlem Center for Complex Quantum Systems, Freie Universit\"at Berlin, Germany}
\orcid{0000-0003-4174-5706}
\email{m.hinsche@fu-berlin.de}

\author{Jonas Haferkamp}
\affiliation{Author list in pseudorandom order. All authors contributed equally.}
\affiliation{Dahlem Center for Complex Quantum Systems, Freie Universit\"at Berlin, Germany}
\affiliation{School of Engineering and Applied Sciences, Harvard University, Cambridge, MA 02318, USA}
\affiliation{Department of Mathematics, Saarland University, Saarbrücken, Germany}

\date{}

\begin{document}

%%%%%%
%
%  TITLE, ABSTRACT & ToC
%
%%%%%%

\maketitle

\begin{abstract}
\noindent In this work, we show that learning the output distributions of brickwork random quantum circuits is average-case hard in the statistical query model. 
This learning model is widely used as an abstract computational model for most generic learning algorithms. 
In particular, for brickwork random quantum circuits on $n$ qubits of depth $d$, we show three main results:
\begin{itemize}
    \item At super logarithmic circuit depth $d=\omega(\log(n))$, any learning algorithm requires super polynomially many queries to achieve a constant probability of success over the randomly drawn instance.
    \item There exists a $d=O(n)$, such that any learning algorithm requires $\Omega(2^n)$ queries to achieve a $\Omega(2^{-n})$  probability of success over the randomly drawn instance.
    \item At infinite  circuit depth $d\to\infty$, any learning algorithm requires $2^{2^{\Omega(n)}}$ many queries to achieve a $2^{-2^{O(n)}}$ probability of success over the randomly drawn instance.
\end{itemize} 
As an auxiliary result of independent interest, we show that the output distribution of a brickwork random quantum circuit is constantly far from any fixed distribution in total variation distance with probability $1-O(2^{-n})$, which confirms a variant of a conjecture by Aaronson and Chen.
\end{abstract}

\newpage

\tableofcontents

%%%%%%
%
%  MAIN TEXT
%
%%%%%%

\section{Introduction}
Quantum circuits are of central importance in quantum computing and serve as a discrete toy model for the physical world. 
Understanding the intrinsic properties of quantum circuits is thus of fundamental interest.
One such property is quantum circuit complexity, which corresponds to the minimum number of elementary gates necessary to implement a given quantum circuit.
Another such property is the complexity of simulating quantum circuits, which is the basis for many quantum advantage proposals.
There, one asks what computational resources are required for sampling from the output distribution of a given quantum circuit when applied to a fixed input product state and when measured in the computational basis. In this work, we take the perspective of  learning theory, and study the complexity of learning the output distribution of quantum circuits. At a high level, this amounts to the resources required to reproduce samples according to the output distribution of a quantum circuit when given black-box access to the corresponding output distribution.

More specifically, we study the \emph{average case} complexity of learning the output distributions of quantum circuits. This amounts to the cost of learning when the quantum circuit is drawn randomly according to some measure. We note that in the setting of quantum circuit complexity the average-case setting has been the subject of intense work, due to connections between randomly drawn quantum circuits and holographic models in high-energy physics \cite{brandao_complexity_2021}. Similarly, 
the established average case hardness of classically simulating quantum circuits has been fundamental to proposals for quantum advantage \cite{hangleiter22}. In our setting, we ask the following question:

\begin{center}\textit{What is the complexity of learning generic quantum circuit output distributions?}\end{center}

We answer this question within the \emph{statistical query} (SQ)  framework by proving lower bounds on the query complexity required to learn only a fraction of
the output distributions of random quantum circuits.
% Apart from being a natural question from the perspective of learning theory, it also has a physical interpretation. 
\ryantext{We note that this problem is the average case version of the natural quantum extension of learning the output distributions of classical circuits~\cite{kearns_efficient_1998}. Additionally, it also has a physical interpretation.}
% On a high level, 
\ryantext{Specifically, on a high level, }the learning problem we consider  corresponds to the setting where an observer living in a world governed by quantum physics is to learn a model of its environment with respect to the outcomes of measurements in a fixed basis. 

\ryantext{We stress that while in this work we consider a ``model'' of the unknown quantum process to be an algorithm for generating new samples from the distribution obtained by measuring the output state in a fixed basis, there are indeed many other notions of both ``model'' and access to the unknown quantum process one could consider. For example, one could allow access to copies of the output state of the quantum circuit, and ask for an algorithm which is only required to provide estimates of expectation values for some fixed set of observables~\cite{arunachalamGuestColumnSurvey2017,anshuSurveyComplexityLearning2024}. Alternatively, one might even only be interested in testing whether the unknown quantum process satisfies a certain property or not (for example, such as being a Clifford circuit). Each such setting will model a different scenario of interest, however in this work we consider the setting of learning a model to generate new samples from fixed basis measurements, in light of both the physical and learning-theoretic interpretations discussed above, as well as a variety of connections to quantum machine learning.}

% We note that this 
% contrasts the setting often considered in 
% quantum learning theory where the learner is granted access to the quantum state on which measurements are made~\cite{arunachalam2017survey}. \ryantext{Additionally, }

% Our results also have implications in the context of quantum machine learning. 
% Firstly, many quantum machine learning proposals today operate within the setting of classical data, and therefore such a restriction is motivated not only from a theoretical but also from a practical perspective. 
% Secondly,
\ryantext{More specifically,} the output distributions of parameterized local quantum circuits \ryantext{with respect to computational basis measurements} are often used as a model class \ryantext{in quantum machine learning approaches to} probabilistic modelling, \ryantext{where they} are referred to as \textit{quantum circuit Born machines} 
(QCBMs). \ryantext{As such, } understanding the extent to which learning algorithms for QCBMs may offer advantages over purely classical approaches, 
such as those based on deep neural networks, is currently the subject of much interest~\cite{hinsche_single_2022,coyle2020born}. 
\ryantext{Our work can be seen as investigating the average-case learnability of the model class of QCBMs, and as such, a rigorous first step towards characterizing the limitations of QCBM based algorithms.}
% By investigating the average-case learnability of the model class of QCBMs, our work can be seen as a rigorous first step towards characterizing the limitations of QCBM based algorithms. 

Additionally, a core technical ingredient of our results has ramifications for quantum advantage proposals based on sampling from the output distributions of random quantum circuits. 
We prove that with overwhelming probability, the output distribution of a random quantum circuit will have at least some constant total-variation distance from the uniform distribution. 
This resembles a conjecture by Aaronson and Chen~\cite{aaronsonComplexityTheoreticFoundationsQuantum2017a} where they conjecture the same statement with different constants. They required and proved a weaker form of this statement for establishing the complexity theoretic foundations of quantum advantage proposals. 
We provide more details on the connections between our work and the variety of related works in Section~\ref{sec:related-work}.

\subsection{Set-up}\label{ss:setup}
\textbf{General framework: }
We say that a class $\D$ of distributions can be learned by an algorithm $\A$ if, when given access to any $P\in\D$, the algorithm returns a description of some close distribution $Q$.
In particular, for an accuracy $\epsilon>0$ we say that $\A$ $\epsilon$-learns $P$ if it returns a description of a $Q$ which is $\epsilon$-close in total variation distance.
An algorithm is said to be a statistical query algorithm if it has access to $P$ only via approximate expectation values instead of individual samples, where the approximation is promised to be within a tolerance $\tau$.
This is not only a handy restriction on the algorithm that makes analysis simpler, but also practically inspired 
 since most heuristic algorithms 
are 
of this form
\cite{feldman2017statistical}. In particular, for $\tau=\Omega(1/\poly(n))$ the statistical query oracle can be simulated from polynomially many samples. 
Formal definitions are given in \Cref{sec:prelim}. 

\textbf{Average case complexity: }
The average case complexity of an algorithm characterizes the cost to achieve a certain probability of success with respect to a measure over the instances\footnote{Thus, we work in the Monte Carlo framework of random algorithms. One can likewise characterize the average case complexity by the \textit{expectation} of success, which then corresponds to the Las Vegas framework. The Monte Carlo framework is more general in our case as the task we consider can in general not be verified efficiently.}.
The (deterministic) average case query complexity with respect to $\mu$ and $\beta$ 
corresponds to the minimal number of queries any algorithm must make in order to have at least a success probability of $\beta$ with respect to $P\sim\mu$.
The randomized average case query complexity is defined in the same manner, though introducing another parameter $\alpha$ capturing the success probability over the internal (classical or quantum) randomness of the algorithm.

\textbf{Quantum and probabilistic algorithms:}\label{intro:random-algorithms}
As explained in more detail in \Cref{app:randomized} the bounds for deterministic algorithms apply almost exactly to random, i.e., probabilistic or quantum, algorithms. 
In particular, the randomized average case complexity for $\epsilon$-learning is lower bounded by the deterministic average case complexity of $\epsilon$-learning up to a prefactor of $2(\alpha-1/2)$ (c.f. \Cref{lem:deterministic-average-complexity-2} in comparison with \Cref{thm:random-average-complexity}), 
where we denote by $\alpha$ the success probability with respect to the internal randomness of the algorithm.
Thus, for the sake of ease of presentation, throughout this work we focus on the deterministic case. We refer to \Cref{app:randomized} for the extension to probabilistic and quantum algorithms.

\textbf{Random quantum circuits: }
The distribution class $\D$ we consider is given by QCBMs, the set of Born distributions corresponding to brickwork quantum circuits at depth $d$ and with gates from a gate set $\G$. In particular, we consider distributions of the form
\begin{align}
    P_U(x)=\abs{\mel{x}{U}{0^n}}^2\,,
\end{align}
with $U$ being some unitary brickwork circuit composed of gates from 
the gate-set $\G$.
While our main focus is 
on $\G=\U(4)$, the set of unitary two qubit gates, our techniques will carry over and give similar results for discrete approximations of $\U(4)$.

Average case bounds and their interpretation depend on the choice of the underlying measure. In this work, we consider the measure over distributions $P_U$ that is induced by sampling a random quantum circuit $U$.
This is, each gate is sampled individually from the the uniform measure over the gate set $\G$.
For $\G=\U(4)$ this measure is well studied and known as the unitary Haar measure.
One practical aspect of this measure is that it corresponds to a natural notion of a generic distribution that would potentially be sampled in the lab by individually sampling each local gate. 
Other interesting measures would be the uniform distribution over the actual distribution class $\D$. However, in case of $\G=\U(4)$, this class is continuous. Hence, defining the uniform measure would require a suitable discretization, e.g., by means of an $\epsilon$-net. 
On the other hand, as we will show, the measure induced by random quantum circuits will be bias free in the sense that for any distribution $P\in\D$, the probability of sampling some distribution close to $P$ is exponentially small.
Thus, similar to the uniform distribution, the measure we consider does not introduce any bias towards any distribution.

\subsection{Our results}\label{ss:our_results}
The main contribution of this work consists in characterizing the average-case complexity of learning the output-distributions of random quantum circuits for different circuit depths. 
Here we provide an informal overview of these results.

Our main focus is on the scaling of the average case complexity with respect to both circuit depth $d$ and the success probability over the randomly drawn instance $\beta$:
\begin{enumerate}
\item \textbf{Circuit depth $d$:} As the circuit depth increases, the expressivity of the set of distributions we consider increases as well. 
At $d=1$ all output distributions of local quantum circuit are product distributions and hence, 
can be learned trivially.
At infinite depth, in contrast, local quantum circuit output distributions can represent any distribution.
Thus, they
are not learnable in the worst case.
As such, the complexity of learning must scale with the circuit depth. 
We we are interested in understanding this dependence.
\item \textbf{Probability over instances $\beta$:} Intuitively the larger $\beta$, the harder we expect the average-case learning task to be. 
Specifically, setting $\beta=1$ recovers the worst-case setting, where we require the learning algorithm to succeed on \textit{all} instances. 
Setting $\beta < 1$ implies that we only require the learning algorithm to succeed on a fraction of instances, which makes the average-case learning problem easier. 
In this work we investigate how small $\beta$ can be made, while still guaranteeing hardness of average-case learning.
\end{enumerate}

In addition to the dependency of average-case query-complexity on $d$ and $\beta$, we also investigate the dependence on both the tolerance of the statistical queries $\tau$ and the desired accuracy $\epsilon$. For ease of presentation, in the informal results below we suppress the dependencies on $\epsilon$ and $\tau$
and focus on the case where $\tau=\Omega(1/\poly(n))$ and $\epsilon$ is a sufficiently small  constant.
We refer to the formal statements for details. 

Finally, we stress again that while the results given below are stated for deterministic algorithms, they immediately translate to both probabilistic and quantum algorithms.
This correspondence is sketched in \Cref{intro:random-algorithms} while the details can be found in \Cref{app:randomized}.
With this in mind, we state the main results of this work as follows:

\begin{inftheorem}\label{infthm:main-result}
Let $\epsilon$ small and $n$ large enough and let $\tau=\Omega(1/\poly(n))$.
Let $\A$ be an algorithm for $\epsilon$-learning the output distributions of brickwork random quantum circuits of depth $d$ from $q$ many $\tau$-accurate statistical queries.
Then it holds
\begin{enumerate}
\item\textbf{Infinite depth:} When $d\rightarrow\infty$, $q=2^{2^{\Omega(n)}}$ queries are necessary for any $\beta>2\exp(-2^{n-2}/9\pi^3)=2^{-2^{\Omega(n)}}$ (c.f. \Cref{thm:query_lower_haar_random}).
\item \textbf{Linear depth:} There is a $d'=O(n)$ such that for any $d\geq d'$, $q = \Omega(2^n)$ queries are necessary for any $\beta > 3200\cdot2^{-n}=O(2^{-n})$ (c.f. \Cref{thm:linear-depth-query-complexity}). 
\item \textbf{Sublinear depth:} for any $c\log n\leq d\leq c(n+\log n)$, $q = 2^{\Omega(d)}=2^{\omega(\log(n))}$ queries are necessary for any $\beta> 4/5+\epsilon+\tau=O(1)$, where $c=1/\log(5/4)$ (c.f. \Cref{thm:query-complexity-sublinear-depth}).
\end{enumerate}
\end{inftheorem}

In particular, we find that the average case problem at superlogarithmic depth is hard with constant probability over the random instance.
Moreover, at linear depth we find hardness with probability exponentially close to one.
At infinite depth, the problem becomes hard with probability double exponentially close to one.

\ryantext{A natural question concerns the tightness of these statistical query (SQ) lower bounds. We note that a crude SQ upper bound for any distribution learning problem is simply given by the cardinality $\mathcal{N}$ of an $\epsilon$-net over the set of distributions to be learned.  This follows from the fact that for any given pair of distributions in the $\eps$-net, there is a corresponding optimal distinguishing query function. Such distinguishing queries can be used in a tournament-style SQ algorithm over $\mathcal{N}-1$ rounds to identify the unknown distribution. For brickwork circuits of depth $d$, one may take $\mathcal{N}$ to be the size of a corresponding $\epsilon$-net over the quantum circuits, giving $ \mathcal{N}= \exp[O(nd \log(nd/\epsilon))]$ \cite{zhaoLearningQuantumStates2024}. This can be compared to the above SQ lower bounds: for instance, for linear depth $d =O(n)$ and constant $\epsilon$, we find an SQ upper bound of $q = \exp[O(n^2 \log(n))]$ which is larger but comparable to our lower bound of $q=\Omega(2^n)$. We leave it as an open problem to provide more tightly matching upper bounds to the lower bounds proved in \cref{infthm:main-result}.
}

As a side result
we show that the output distribution of a random quantum circuit is, with overwhelming probability, at least constantly far in total variation distance from any fixed distribution. This resolves a variant of Aaronson and Chen's~\cite[Conjecture 1]{aaronsonComplexityTheoreticFoundationsQuantum2017a}, made with the goal of clarifying the hardness of verifying random circuit sampling procedures.
As such we believe this auxiliary result to be of independent interest.

\begin{inftheorem}[Informal version of \Cref{thm:farfromeverything}]
\label{ithm:farfromeverything}
There exists some $d'=O(n)$ such that for any depth $d\geq d'$, for any \ryantext{$\epsilon\leq {1}/{225}$} and for any distribution $Q$ we have
\begin{align}
\Pr_{U\sim\mu}\lrq{\tv(P_U,Q)> \epsilon} \geq 1-O\left( 2^{-n} \right)\,,
\end{align}
where $\mu$ is the measure induced by random brickwork quantum circuits.
\end{inftheorem}

\subsection{Related work}\label{sec:related-work}
Our work touches on and combines a variety of well studied fields. 
In order to provide further context and motivation, we provide below a discussion of relevant related work.

\textbf{Statistical queries:} 
We work in the 
statistical query (SQ) framework which was introduced by Kearns \cite{kearns_efficient_1998} as a restriction of Valiants theory of learning~\cite{valiant_theory_1984}. Kearns original motivation was the intrinsic robustness of SQ learners with respect to random classification noise. However, the SQ model is also highly relevant in the context of statistical problems~\cite{feldman2017statistical}, which includes distribution learning, as originally formulated by Kearns et al.~\cite{Kearns:1994:LDD:195058.195155}. 
In this context, SQ algorithms are those which only have access to coarse statistical properties of the data generating distribution.
While this is a restriction of the oracle access  
it turns out, with the famous exception of Gaussian elimination, that almost all known learning algorithms can be recast as SQ algorithm \cite{kearns_efficient_1998, feldman_general_2017}.
 
The SQ framework is particularly interesting in the context of variational quantum machine learning, such as QCBM based algorithms. 
To see this we note that all current methods for optimizing the QCBM's parameters use noisy evaluations of gradients, or gradient-like quantities, along individual directions. 
Examples for this include stochastic gradient descent via the parameter shift rule~\cite{mitarai_quantum_2018, schuld_evaluating_2019} and simultaneous perturbations and stochastic approximation~\cite{spall_overview_1998}.
Thus, the update in QCBM based algorithms can be directly implemented via statistical queries. 
\ryantext{The SQ framework has also been used to model variational quantum algorithms beyond QCBMs and prove rigorous lower bounds for them \cite{Anschuetz_2022,nietnerUnifying2023}.}

Equivalences to other statistical oracles, such as the ``honest SQ'' oracle and the statistical query oracle with respect to Bernoulli noise are worked out in \cite{feldman2017statistical, feldman_general_2017}.
Interestingly, SQ learning has been shown to be equivalent to learning with restricted memory \cite{steinhardt_memory_2016, feldman_complexity_2018}, as well as differentially private learning \cite{dworkCalibratingNoiseSensitivity2006, kasiviswanathanWhatCanWe2011}. 
Evolutionary algorithms can be recast as SQ algorithms and in fact were shown to be equivalent to ``correlational statistical query'' (CSQ) algorithms  \cite{valiantEvolvability2009, feldmanEvolvabilityLearningAlgorithms2008a}.
A particularly nice feature of the SQ framework is that it allows for unconditional lower bounds.
As such, SQ lower bounds are often taken as evidence for computational hardness if the underlying problem does not admit a linear structure, as is the case for parities.
Additionally, many statistical query lower bounds asymptotically match complexity theoretic upper bounds, such as learning DNFs \cite{blum1994weakly}, learning mixtures of Gaussians \cite{diakonikolasStatisticalQueryLower2017} and the planted clique problem \cite{feldman2017statistical}.
Other results via the SQ framework contain positive results for k-means clustering, principle component analysis and the perceptron algorithm
\cite{blumPracticalPrivacySuLQ2005a} and manifold estimation \cite{aamariStatisticalQueryComplexity2021}, negative results for learning simple neural networks \cite{chen2022learning}, average case hardness for learning neural networks at super-logarithmic depth \cite{agarwalDeepConditioningTreatment2021}, as well as lower and upper bounds for optimization and distributional search problems \cite{feldman_stat_optimization_2016, feldman2017statistical, feldman_complexity_2018}.

\ryantext{
The SQ framework for learning classical objects such as probability distributions as covered in this work has recently been generalized to the quantum statistical query (QSQ) framework for learning quantum objects such as quantum states or processes \cite{arunachalam2020quantumstatisticalquerylearning,arunchalamRoleEntanglement2023, nietnerUnifying2023, Wadhwa2025learningquantum}. Recently \cite{nietnerUnifying2023}
introduced the evaluation query oracle as an abstraction to unify learning models, which share the key properties of statistical queries, including QSQs, CSQs and parametrized learning algorithms.}

\textbf{Complexity of quantum circuits and computational learning theory:}
A circuit class is a collection of quantum circuits. A well-established way to assess the complexity of such a circuit class from the viewpoint of classical computing is to study the resources required to classically simulate the circuit class. Examples of circuit classes that have been studied in this regard include Clifford circuits, Clifford+$T$ circuits, matchgate circuits and IQP circuits \cite{gottesmanHeisenbergRepresentationQuantum1998, aaronson_gottesman_2004, valiantQuantumCircuitsThat2012, terhalSimulationFermion2002a, bremner11, bremnerAchievingQuantumSupremacy2017}.
Note that in all these examples the description of the circuit class includes a specification of the input state and a fixed measurement basis as the simulatability may crucially depend on these choices.
In this work, we analogously assess the complexity of a quantum circuit class from the viewpoint of computational learning theory. 
Indeed, a long line of research has been aimed at characterizing the complexity of learning \textit{classical} circuit classes, in various learning models~\cite{linial1993constant, kharitonov1993cryptographic, Kearns:1994:LDD:195058.195155, arunachalam2021quantum}. 
This complexity is indeed a fundamental property of such circuit classes. Our work provides insight into this fundamental property of quantum circuits.
We stress here that, analogous to the case of classical simulation, we consider learning with respect to a fixed input state and fixed measurement basis. In that sense, our learning model differs from those employed in other works on learning quantum states \cite{aaronsonLearnabilityQuantumStates2007,montanaroLearningStabilizerStates2017} as we require learning the action of the circuit only with respects to measurements in a fixed product basis.
\ryantext{This is the key difference between the learning task considered in our work and recent work on learning shallow quantum circuits or their output states \cite{huangLearningShallowQuantum2024,zhaoLearningQuantumStates2024,landauLearningQuantumStates2025}, respectively.} 

\textbf{Heuristic algorithms for distribution learning:} Recent years have seen major advances in the development of heuristic neural-network based methods for probabilistic modelling. Generative adversarial networks~\cite{goodfellow2014generative} and generative transformer models~\cite{brown2020language} have managed to achieve impressive results ranging from predicting protein structure to atomic accuracy~\cite{alphafold21} to achieving human-level language comprehension~\cite{Chinchilla22}. 
Learning the underlying model classes is known to be worst case hard~\cite{chen2022learning}. 
This inspired a series of works in order to better understand these successes from a theoretical perspective (see for example \cite{arora_provable_2013, livni_computational_2014, choromanskaLossSurfacesMultilayer2015a, janzaminBeatingPerilsNonConvexity2016, danielySGDLearnsConjugate2017, shamirDistributionSpecificHardnessLearning2017, abbe_poly-time_2020, danielyHardnessLearningNeural2020, agarwalDeepConditioningTreatment2021}). 
Our results can be seen in a similar light for QCBM based algorithms. In particular, our average case hardness results for learning super logarithmic depth quantum circuit distributions is the QCBM equivalent to  \cite[Contribution 3]{agarwalDeepConditioningTreatment2021}.

\textbf{``Far from uniform'' property and quantum advantage:} 
In \cite{aaronsonComplexityTheoreticFoundationsQuantum2017a}, the authors propose \textit{heavy output generation} as a particularly natural task for separating classical and quantum computers.
More precisely, quantum computers can be used to output sets of bit strings $z_1,\ldots,z_k$ such that more than $2/3$ of them have probability larger than the median of the output probabilities.
This can be achieved by a subroutine, which uses a conditional probability distribution that samples instances of random quantum circuits as long as necessary for a sufficiently non-uniform probability distribution to appear. 

A key result is that far-from-uniform output distributions are not rare for random quantum circuits. 
We adapted the resulting bound as \Cref{cor:const_far_from_uniform}.
Aaronson and Chen moreover conjectured that this far from uniform conjecture not only holds with constant probability but with probability exponentially close to one.
This would imply that it suffices to directly sample from a random quantum circuit instance without the subroutine.
While we prove that a far from uniform property holds with exponential probability in \Cref{thm:ball_U}, the constants in our result do not imply Conjecture~1 in \cite{aaronsonComplexityTheoreticFoundationsQuantum2017a}.
At the same time, it is possible to define a weaker version of heavy output generation, for which our bounds suffice.

\textbf{From average-case complexity in learning to cryptography:}
There is a rich correspondence between computational learning theory and cryptography. On the one hand, cryptographic assumptions are often used to prove conditional lower bounds for learning problems~\cite{kearns1994introduction}. On the other hand, the assumed hardness of learning problems can sometimes be used for the construction of cryptographic primitives. For this latter direction, it is well known that the existence of cryptographic primitives such as one-way functions require the existence of learning problems which are \textit{average-case} hard~\cite{IL90,Barak}. As concrete examples, Blum, Furst, Kearns and Lipton~\cite{BFKL} have shown that an efficient average case learner for polynomial size circuits in the distribution specific PAC model, would imply the non-existence of one-way functions. This result has been recently extended by Nanashima \cite{nanashima_theory_2021} to give a characterization of \ryantext{auxiliary}-input one way functions, based on the hardness of PAC learning polynomial size circuits in a modified average-case variant of PAC learning. In light of these results, it is natural to ask whether one can characterize either classical or quantum cryptographic primitives in terms of the complexity of average-case learners for \textit{quantum} circuit output distributions in the distribution learning setting. While we do not address this question in this work, and while our restriction to the SQ model is a major restriction in this regard, we believe that 
the connection to cryptography
merits further study and hope that the insights gained from our results can be helpful in this regard.

\subsection{Proof overview}\label{ss:proof_over}

For ease of presentation, we use the notation $P[\phi]$ to denote $\Ex_{x\sim P}[\phi(x)]$  and denote by $\mc U$ the uniform distribution. 
The starting point of all our results is a lower bound on the average case query complexity in terms of  properties of the measure $\mu$.  
Suppose there is a deterministic algorithm $\A$ that $\epsilon$-learns a $\beta$ fraction of $\D$ with respect to $\mu$ from $q$ many $\tau$-accurate statistical queries.
Then, it holds (c.f. \Cref{lem:deterministic-average-complexity-2})
\begin{align}\label{eq:proofover-deterministic-average-complexity-2}
q+1\geq\frac{\beta-\Pr_{P\sim\mu}\lrq{\tv(P,\mc U)\leq\epsilon+\tau}}{\max_\phi\Pr_{P\sim\mu}\lrq{\abs{P[\phi]-\mc U[\phi]}>\tau}}\,,
\end{align}
where the $\max$ is over all bounded functions $\abs{\phi(x)}\leq1$.
 
The above bound is obtained by first reducing a suitable worst-case uniformity test to the average-case learning problem, and then lower bounding the complexity of the uniformity test. Given some set of distributions $\widetilde\D\subseteq\D$, we consider the decision problem of testing $\widetilde{\D}$ versus $\mc U$ defined via:

\begin{center}
\textit{Given statistical query access to some $P\in \widetilde{\D}\cup \lrb{\mc U}$ decide whether ``$P=\mc U$'' or ``$P\in\widetilde{\D}$''.}
\end{center} 

We then note:

\begin{enumerate}
\item An \textit{average-case} learner for $\D$ which succeeds on a $\beta$ fraction of instances with respect to $\mu$, implies the existence of a \textit{worst-case} learning algorithm for some $\D'\subseteq \D$ with $\mu(\D') =\beta$. This trivially implies a worst-case learning algorithm for $\widetilde{\D} = \D'/B_{\epsilon + \tau}(\mc U)$.
\item A \textit{worst-case} learning algorithm for $\widetilde{\D}$ using at most $q$ queries implies an algorithm for deciding $\widetilde{\D}$ versus $\mc U$ using $q+1$ queries. 
\end{enumerate}
As such, it is sufficient for us to lower bound the complexity the uniformity test. To this end, we use a counting argument to show that for any measure $\nu$ over $\widetilde{\D}$ it holds that the number of queries necessary to decide $\widetilde{\D}$ versus $\mc U$ satisfies
\begin{align}\label{eq:sketch-deciding-complexity}
        q\geq\lr{\max_\phi\Pr_{P\sim\nu}\lrq{\abs{P[\phi]-\mc U[\phi]}>\tau}}^{-1}\,.
\end{align}
We then obtain \Cref{eq:proofover-deterministic-average-complexity-2} by considering the measure $\nu$ defined by conditioning $\mu$ on $\widetilde{\D}$.

Given this, from \Cref{eq:proofover-deterministic-average-complexity-2} it is clear that in our context, in order to obtain the desired lower bounds we require:
\begin{itemize}
    \item An upper bound on the maximal fraction distinguishable from uniform $\f=\max_{\phi}\Pr_U[ |P_U[\phi_i]-\mathcal{U}[\phi_i]|\geq \tau]$.
    \item An upper bound on the mass of the $\epsilon$-ball around the reference distribution. The complement of this probability is often referred to as the probability of being \textit{far from uniform}.
\end{itemize}

In this work we give bounds on both quantities for random quantum circuits of various depths. We begin in the limit of infinitely deep circuits, and thus of Haar-random unitaries and then partially derandomize our results using concentration inequalities based on higher moments of the Haar measure. This will allow us to prove results about random quantum circuits of comparably low depth, which are far from the Haar measure but quickly generate the same moments.

\subsubsection{Bounding $\f$}

For Haar random unitaries, the concentration of measure phenomenon in the form of Levy's lemma produces tight bounds on tails, which implies bounds on $\f$.

For linear depth circuits we can use Chebyshev's inequality in order to bound $\f$ 
\begin{align}
    \max_{\phi}\Pr_U[ |P_U[\phi_i]-\mathcal{U}[\phi_i]|\geq \tau]\leq\frac{\Var_{U}\lrq{P_U[\phi]}}{\tau^2}
\end{align}
where we have used $\Ex_{U}\lrq{P_U[\phi]}=\mc U[\phi]$ since for any depth $d\geq1$ $\mu$ is a 1-design.
The variance is a second moment and can be exactly computed for the Haar measure via standard symmetry arguments or from the Weingarten calculus.

For sublinear circuits, the distribution over unitaries does not form a unitary design, however, we can still bound the second moments involved in the variance by adapting the statistical physics mapping from \cite{hunterjones19} in a similar way as in \cite{barak2021spoofing}.
Many of our bounds will depend on such moment bounds over the Haar measure that we detail in \Cref{app:moments}.

\subsubsection{Bounding the far from uniform probability}\label{sss:munought} 
To obtain a lower bound on the far from uniform probability (or equivalently an upper bound on the probability mass of an $\epsilon$-ball around the uniform distribution), we start by writing the total variation distance in terms of the $\ell_1$-norm. 

In the infinite circuit depth regime we make use of Gaussian integration in order to obtain upper and lower bounds on the expected distance between a random output distribution $P_U$ and the uniform distribution. Then we use Levy's Lemma once again to obtain a bound on the probability of $P_U$ being $\epsilon$ far from the uniform distribution.

In the linear depth regime we apply a variant of Berger's inequality for $\ell_p$-norms, c.f. \Cref{lemma:berger_for_norms}, in order to obtain
\begin{align}
   \Pr_{U}\lrq{\norm{P_U-\mc{U}}_1<2\eps}
    \leq\Pr_{U}\lrq{\frac{\norm{P_U-\mc{U}}_2^3}{\norm{P_U-\mc{U}}_4^2}<2\eps}\,.
\end{align}
Treating the numerator and denominator separately we can thus apply the union bound and obtain 
\begin{align}
     \Pr_{U}\lrq{\norm{P_U-\mc{U}}_1<2\eps}\leq\Pr_{U}\lrq{\norm{P_U-\mc{U}}_2^3<2\eps_1}+\Pr_{U}\lrq{\norm{P_U-\mc{U}}_4^2>2\eps_2}\,,
\end{align}
for suitable $\frac{\eps_1}{\eps_2}\geq\eps$. Both terms can  be bounded separately with the same strategy. 
Using Chebyshev's inequality for the random variable $X(p,q)=\norm{P_U-\mc{U}}_p^q$ we can bound the deviation of $X(p,q)$ from its mean $\Ex\lrq{X(p,q)}$. In particular, due to the variance term in Chebyshev's inequality we find that an (approximate) $2p$-design is sufficient for an exponential concentration of $X(p,q)$. Since $\Ex[X(2,3)]$ is exponentially small it is thus crucial, that $X(4,2)$ itself is sufficiently small and sharply concentrated such that we can find $\epsilon_1$ and $\epsilon_2$ that cancel to a constant. Again, we confirm this ingredient via a variance bound provided that our measure is induced by an $8$-design. We conclude from \cite{brandao2016local, haferkamp2022random} that random linear depth Born distributions are far from uniform.

In the sublinear depth regime we use a result by Aaronson and Chen \cite{aaronsonComplexityTheoreticFoundationsQuantum2017a} which lower bounds the expected distance between a randomly drawn $P_U$ and the uniform distribution even for $d=1$. We then use Markov's inequality to translate this to a bound the probability of $P_U$ being $\epsilon$ far from the uniform distribution.

\subsection{Discussion and future work}\label{ss:discussion}

In this work we give lower bounds for the average case query complexity of learning the output distributions of random quantum circuits in different depth regimes. 
In particular, we show that the problem of learning the output distribution of random quantum circuits is hard with constant probability over the instance already at super logarithmic depth. 
Moreover, we prove that the problem becomes hard with probability exponentially close to one over the instance at linear depth.
Our analysis is accompanied by the corresponding results for both Haar random unitary output distributions and Haar random Clifford output distributions. 
While the former gives hardness with probability doubly exponentially close to 1, the latter only gives hardness with a constant probability over the instance. 

There are multiple natural avenues to continue this work:

\begin{enumerate}
\item While we prove a strong asymptotic average-case complexity bound for linear depth circuits, the explicit depth at which our bounds apply comes with a rather large prefactor ($d=10^{20}n$).
This prefactor is likely an artefact of the proof techniques in \cite{brandao2016local,haferkamp2022random}, which bound the depth at which unitary designs are well-approximated.
There are at least two promising approaches to lowering this constant and consequently making our bound more directly applicable to a \ryantext{practical} regime.
One could directly compute the moments using the statistical physics mapping from \cite{hunterjones19}.
Alternatively, extensive numerical calculations of finite-size spectral gaps in combination with Knabe bounds might lower the explicit depth at which unitary designs are generated~\cite{haferkamp2021improved}.
While these calculations are beyond the scope of this paper, we believe that it would be worthwhile to address this issue.

\item Moreover, when relaxing the assumption on $\epsilon$ from constant to inverse polynomial and on $\beta$ from inverse exponential to inverse polynomial, thus making the learning task harder, one can use Markov's inequality instead of Chebyshev's in the far from uniform proof (c.f. \Cref{thm:4-design}). 
This already improves the asymptotics to hold at $d=690n$.  
This might simplify the corresponding statistical physics mapping since it makes use of fourth instead of eighth moments. 
We expect that such a calculation implies average-case hardness with probability $1-o(1)$ over the instances for any depth $d=\omega(\log(n))$.

\item In this work, we rule out efficient algorithms at super logarithmic depth. 
It is thus natural to ask what happens at logarithmic depth and below. 
For example, are the output distributions of constant depth quantum circuits efficient to learn? 

\item Last, we emphasize that all our results hold in the statistical query model. 
Another prominent model is learning from samples.
With the famous exception of parities, most known tight upper bounds for learning can be realized in the statistical query model.
We ask whether our average-case complexity results carry over to average-case hardness of learning from samples. 
\end{enumerate}

\section{Notation and preliminaries}\label{sec:prelim}

\subsection{Statistical query learning}

Let $\D$ be a class of distributions over a domain $X$. For two distributions $P,Q\in\D$  we denote by ${\tv(P,Q):= \frac{1}{2} \sum_{x\in X} |P(x)-Q(x)|}$ the total variation distance between them. 
The open $\eps$-ball $B_\eps(P)$ around any distribution $P$ over the domain $X$ is given by the set of all distributions $Q$ over $X$ such that $\tv(P,Q)<\eps$.
For a distribution $P$ over $X$ and a function $\phi:X\rightarrow[-1,1]$ we use the short hand notation
\begin{align}
    P[\phi]\defeq\Ex_{x\sim P}[\phi(x)]
\end{align}
to refer to the expectation value of $\phi$ with respect to $P$.
We denote by $\D_X$ the set of distributions over $X$ and by $\Dn$ the set of all distributions over the domain $\01^n$. The uniform distribution is denoted by $\mc U$.

A well studied model in learning theory is the statistical query learning model. Here we assume that the learner has access to expectation values of functions with respect to the underlying probability distribution. This can be formalized by considering access to a statistical query oracle. 

\begin{definition}[Statistical query oracle]\label{def:stat}
For $\tau>0$ and a distribution $P$ over $X$ we denote by $\stat_\tau(P)$ the statistical query (SQ) oracle of $P$ with tolerance $\tau$. When queried with some function $\phi:X\to[-1,1]$ the oracle returns some $v$ such that $\abs{v-P[\phi]}\leq\tau$.
\end{definition}

\begin{remark}
On immediate consequence of \Cref{def:stat} which is useful for the interpretation of our formal result statements is as follows: 
For any $\zeta<\tau$, any SQ oracle $\stat_{\zeta}(P)$ is also a valid SQ oracle $\stat_\tau(P)$. 
Thus, lower bounds for SQ algorithms with respect to $\zeta$ trivially imply the same lower bound for SQ algorithms with respect to $\tau$.
\end{remark}

A prominent special case is to consider $\tau$ to be lower bounded by an inverse polynomial, since this reflects the scenario where the statistical query oracle can be run efficiently with a polynomial number of samples. 
Up to polynomial corrections, the complexity of the oracle is then given by the complexity of computing the query function $\phi$.

In order to learn a distribution it is crucial to fix the representation of the distribution to be learned.
Here, a representation can be thought of as an algorithm that specifies the distribution. 
For the sake of clarity typical representations are generators and evaluators. 
\begin{itemize}
    \item A generator of a probability distribution $P$ is a probabilistic algorithm that produces samples according to $x\sim P$.
    \item An evaluator of a probability distribution $P\in\Dn$ is an algorithm $\eval_P:\01^n\rightarrow[0,1]$ that outputs the probability amplitude $\eval_P[x]=P(x)$.
\end{itemize}
We will refer to a representation (e.g. a generator  or an evaluator) of a distribution $Q$ as an $\epsilon$-approximate representation of $P$ if $\tv(P,Q)<\epsilon$. 
We would like to stress that, due to fundamental limitations, our results hold for any $\epsilon$-approximate representation even if we allow the corresponding algorithms to be computationally inefficient.

Distribution learning in the statistical query framework is made formal by the following definition.

\begin{problem}[$\eps$-learning of $\D$ from statistical queries]\label{prob:learning} 
Let $\eps\in(0,1)$ be an accuracy parameter, $\tau\in(0,1)$ be the tolerance and let $\D$ be a distribution class. For a fixed representation of the distribution, the task of $\eps$-learning $\D$ from statistical queries with tolerance $\tau$ is defined as given access to $\stat_\tau$ for any unknown $P\in\D$, output an $\eps$-representation of $P$.
\end{problem}

In \cite{hinsche_single_2022} the authors have studied the worst case query complexity of \Cref{prob:learning} for $\D$ being the class of output distributions of local quantum circuits. Here we want to consider the average case query complexity for the same distribution class. This is a strictly easier task as we do not require the learner to succeed for each and every distribution in the class. 
Rather, we want to characterize the number of statistical queries needed to succeed in solving 
\Cref{prob:learning} on a fraction of distributions in $\D$ with respect to a measure $\mu$ over the distributions in $\D$. 
Similarly, when one considers a quantum or probabilistic learner one  can ask about the number of statistical queries needed to succeed in the aforementioned task with some fixed probability with respect to the algorithm's randomness.
Thus, the randomized average case complexity is defined with respect to a measure $\mu$ and the two parameters $\alpha$ and $\beta$. By $\alpha$ we denote the success probability of the algorithm and by $\beta$ the size, with respect to $\mu$, of the fraction on which the algorithm is successful.

\begin{definition}[Average case complexity]\label{def:averagecasecomplexity}
Let $\D$ be a class of distributions, $\mu$ a probability measure over $\D$ and $\alpha,\beta\in(0,1)$.
The deterministic average case query complexity of \Cref{prob:learning} is defined as the minimal number $q$ of queries any learning algorithm $\A$ must make in order to achieve
\begin{align}
    \Pr_{P\sim\mu}\lrq{\text{``$\A^{\stat(P)}$ $\epsilon$-learns $P$ from $q$ queries''}}\geq\beta\,.
\end{align}
Likewise, the randomized average case query complexity is defined as the minimal number $q$ of queries any random learning algorithm $\A$ must make in order to achieve
\begin{align}
    \Pr_{P\sim\mu}\lrq{\Pr_{\A}\lrq{\text{``$\A^{\stat(P)}$ $\epsilon$-learns $P$ from $q$ queries''}}\geq\beta}\geq\alpha\,,
\end{align}
where $\Pr_\A$ denotes the probability over the internal randomness of $\A$.
\end{definition}

The deterministic and randomized average case complexities in the framework of statistical query learning are closely related. 
As advertised in \Cref{intro:random-algorithms}, one can directly translate our lower bounds for deterministic learning to lower bounds of randomized learning by means of a global prefactor $2(\alpha-1/2)$.
Thus, for the sake of ease we focus on the deterministic average case complexity throughout the main text and refer to \Cref{app:randomized} for the details about random algorithms. 

As discussed in \Cref{ss:proof_over} the average case query complexity of deterministic algorithms for learning in the SQ framework can now be lower bounded as follows.

\begin{lemma}[Deterministic average case complexity]\label{lem:deterministic-average-complexity-2}
Suppose there is a deterministic algorithm $\A$ that $\epsilon$-learns a $\beta$ fraction of  $\D$ with respect to $\mu$ from $q$ many $\tau$-accurate statistical queries. Then for any $Q$ it holds
\begin{align}\label{eq:lem-deterministic-average-complexity-2}
q+1\geq\frac{\beta-\Pr_{P\sim\mu}\lrq{\tv(P,Q)\leq\epsilon+\tau}}{\max_\phi\Pr_{P\sim\mu}\lrq{\abs{P[\phi]-Q[\phi]}>\tau}}\,,
\end{align}
where again, the $\max$ is over all functions $\phi:X\to[-1,1]$.
\end{lemma}

We refer to \Cref{app:omitted-proof-lemma} for the proof of \Cref{lem:deterministic-average-complexity-2}. To provide a simplified expression we make the following remark.

\begin{remark}\label{rem:lem-average-case-complexity}
Note that without loss of generality we can take $\tau \leq \epsilon$ which leads to the bound
\begin{equation}
    q+1\geq\frac{\beta-\Pr_{P\sim\mu}\lrq{\tv(P,Q)\leq 2\epsilon}}{\max_\phi\Pr_{P\sim\mu}\lrq{\abs{P[\phi]-Q[\phi]}>\tau}}.
\end{equation}
To see why we can do so, consider instead the case $\tau > \epsilon$. Given $P,Q$ such that $\tau>\tv(P,Q)>\epsilon$ we can see that these distributions are indistinguishable with respect to $\tau$-accurate queries and thus there cannot exist an $\epsilon$-learner.
\end{remark}

From \Cref{lem:deterministic-average-complexity-2} it is clear that a crucial figure of merit is the fraction of distributions that can be distinguished from a single query. 
Following \cite{feldman_general_2017} we define.

\begin{definition}[Maximally distinguishable fraction]\label{def:frac}
Let $\D$ be a distribution class over the domain $X$ and let $\mu$ 
be some probability measure over $\D$. The maximally distinguishable fraction with tolerance parameter $\tau$ and with respect to the measure $\mu$ and the reference distribution $Q$ is defined as
\begin{align}
    \kfrac(\mu, Q, \tau) \defeq\max_{\phi}\Pr_{P\sim\mu}\lrq{\abs{P[\phi]-Q[\phi]}>\tau}\,,
\end{align}
where the maximum is over all functions $\phi:X\rightarrow[-1,1]$. 
\end{definition}

In the special case that the reference distribution is the uniform distribution, as is the case in the remainder of this paper, $\mc U$ we will refer to this by the short hand
\begin{align}
    \f=\kfrac(\mu, \mc U, \tau)\,,
\end{align}
where the measure $\mu$  and the tolerance $\tau$ will be clear from context.

\textbf{Summary:} A lower bound for average case query complexity for learning in the statistical query model is determined by: 
\begin{itemize}
    \item The size of the $\epsilon+\tau$-ball of any fixed reference distribution $\Pr_{P\sim\mu}[\tv(P,Q)\leq\epsilon+\tau]$.
    \item The maximally distinguishable fraction with respect to the same reference distribution $\kfrac(\mu,Q,\tau)$.
\end{itemize}

\begin{note}
For the sake of ease of presentation we give the derivation of bounds on the weights of the ball around the reference distribution in terms of $\epsilon$. 
We then translate the corresponding result to  \Cref{lem:deterministic-average-complexity-2}  substituting $\epsilon$ by $\epsilon+\tau$.
\end{note}

\subsection{Random quantum circuits}
Given some $n$-qubit unitary $U\in\U(2^n)$, we denote by $P_U(x) = \abs{\mel{x}{U}{0^{n}}}^2$ the quantum circuit output, or Born distribution.
We denote by $\muU$ the  unitary Haar measure, or simply the  uniform measure, over $\U(D)$. Similarly, we denote by $\muS$ the spherical Haar measure, or likewise the uniform measure, over the complex unit sphere $\mb S^{D-1}$,  where in both cases the dimensionality $D$ will be clear from context. 

\begin{definition}[Brickwork architecture]\label{def:brickwork}
An $n$-qubit brickwork quantum circuit of depth $d$ (with periodic boundary conditions) is a quantum circuit that is of the form
\begin{align}
\begin{split}
    U=
    (U_{2,3}^{(d)}\otimes\cdots\otimes U_{n,1}^{(d)})
    &\cdot
    (U_{1,2}^{(d-1)}\otimes\cdots\otimes U_{n-1,n}^{(d-1)})
    \cdots\\[5pt]
    &\cdots 
    (U_{2,3}^{(2)}\otimes\cdots\otimes U_{n,1}^{(2)})
    \cdot
    (U_{1,2}^{(1)}\otimes\cdots\otimes U_{n-1,n}^{(1)})
\end{split}
\end{align}
where $U_{i,j}^{(k)}\in\U(4)$ is the unitary in the $k$'th layer acting on neighboring  qubits $i$ and $j$. 
For the sake of ease we have assumed $d$ and $n$ to be even.
\end{definition}

While we give definitions and analysis only for periodic boundary conditions, we note that all our results will carry over to open boundary conditions at the price of slightly worse prefactors. 

\begin{definition}[Random quantum circuits]\label{def:rqc}
A random brickwork quantum circuit of depth $d$ on $n$ qubits is formed by drawing
$\lfloor n/2\rfloor\cdot d$ many 2-qubit unitaries $U_{i,j}^{(k)}$ i.i.d. Haar randomly and contracting them.
We denote the resulting probability distribution on $\U(2^n)$ by $\muC$, where $n$ and $d$ will be clear from context.
\end{definition}

Given a two-qubit gate set $\G\subseteq\U(4)$. We denote by $\D_\G(n,d)$ the set of Born distributions which can be realized by brickwork quantum circuits on $n$ qubits of depth $d$.

\subsection{Unitary designs}
Unitaries generated by random quantum circuits quickly mimic Haar random unitaries for many practical purposes. 
The reason for this is that they generate unitary $t$-designs.
These are "evenly" spread probability distributions over the unitary group that have the same $t$'th moments as the Haar measure~\cite{dankert2005efficient,gross2007evenly}.
This is often expressed in terms of $t$-fold twirls: 
Let $\nu$ be a probability measure on the unitary group $\U(D)$. Then we define for any matrix $A\in \mb{C}^{D^t\times D^t}$.
\begin{equation}
    \Phi^{(t)}(\nu)(A):=\int U^{\otimes t} A (U^{\dagger})^{\otimes t}\mathrm{d}\nu(U)\,.
\end{equation}
We call $\nu$ an approximate unitary $t$-design if, for $\muU$ being the Haar measure.
\begin{equation}
    \Phi^{(t)}(\nu)\approx\Phi^{(t)}(\muU)\,.
\end{equation}

We provide a detailed definition in Appendix~\ref{appendix:unitarydesigns}.
Moreover, see Appendix~\ref{appendix:unitarydesigns} for the relation to state designs and bounds on the generation of approximate designs by random quantum circuits.

\section{Haar random unitaries}

In this section, we bound the two key quantities, namely $\mathfrak{f}$ and the far from uniform probability, for random quantum circuits of infinite depth, corresponding to Haar random unitaries. 
Plugging these bounds into \Cref{lem:deterministic-average-complexity-2}, we obtain the following lower bound on the average case query complexity.

\begin{theorem}[Formal version of infinite depth part of \Cref{infthm:main-result}]\label{thm:query_lower_haar_random}
Let $\tau>0$, $\epsilon\leq 1/e - 2^{-n/2-1}-\tau$ and set $\xi = 1/e -  2^{-n/2-1} - \epsilon-\tau$. Any algorithm that succeeds in $\epsilon$-learning a $\beta$ fraction of the output distributions of infinitely deep random brickwork quantum circuits requires $q$ many $\tau$-accurate statistical queries, with
\begin{equation}\label{eq:query_complexity_infinte}
q+1\geq \frac{\beta- 2 \exp \left( - \frac{2^{n+2} \xi^2}{9 \pi^3} \right)}{2 \exp \left( - \frac{2^{n} \tau^2}{9 \pi^3}\right)}\,.
\end{equation}
\end{theorem}

\begin{remark}
Note that, for any  
\begin{align*}
    \tau\geq2^{-n/4}\quad\text{and any}\quad \epsilon\leq\frac1e-2^{-n/2-1}-2^{-n/4+2} - \ryantext{\tau}\,
\end{align*}
which corresponds to $\xi\geq2^{-n/4+2}$, we find by \Cref{thm:query_lower_haar_random} the query complexity for learning any fraction $\beta>2\exp(-2^{n/2+4}/9\pi^3)=2^{-2^{\Omega(n)}}$ requires $q=2^{2^{\Omega(n)}}$ many queries.
 In words: learning a doubly exponentially small fraction takes doubly exponentially many, inverse exponentially accurate statistical queries.
\end{remark}

 In the case of infinitely deep circuits, 
 it is known that the distribution of circuit unitaries converges to the unique, rotationally invariant Haar measure on the full unitary group in $D=2^n$ dimensions $\muU$. 
If we apply such a Haar-random unitary to a designated (arbitrary) pure starting state, say $|\psi_0 \rangle = |0,\ldots,0\rangle$, we obtain a pure state $\ket{\psi}$ that is sampled from the spherical Haar measure $\muS$, i.e. uniformly from the set of all pure states in $D$ dimensions:
\begin{equation*}
\ket{\psi} = U |\psi_0 \rangle \overset{\textit{unif}}{\sim} \left\{ |u \rangle \in \mathbb{C}^D:\; \langle u|u \rangle =1 \right\} \subset \mathbb{C}^{D} \quad \text{(and $D=2^n$ for $n$ qubits)}.
\end{equation*}
Such (Haar) uniform distributions of pure states have two remarkable features: (i) we can use powerful frameworks like Weingarten calculus and Gaussian integration to compute complicated expectation values and (ii) concentration of measure (Levy's lemma) asserts that concrete realizations concentrate very sharply around this expected behavior. 
These two features can be combined into a powerful strategy to obtain very sharp bounds on deviation probabilities, like the two essential ingredients in \Cref{lem:deterministic-average-complexity-2}:
\begin{equation*}
\Pr_{U \sim \muU}[\tv(P_U, \mathcal{U})\geq \epsilon] \quad \text{and} \quad
\Pr_{U \sim \muU}[\abs{P_U[\phi]-\mathcal{U}[\phi]}>\tau]
\end{equation*}
for any fixed function $\phi: \left\{0,1\right\}^n \to \left[-1,1\right]$.
To apply this formalism, our strategy will be to first reformulate the arguments in both probabilities as functions in the (Haar) random pure state $\ket{\psi} = U|\psi_0 \rangle$. Then we will use Levy's lemma to obtain bounds on both probabilities. Let's focus on this last step as it applies to both probabilities.

Levy's lemma asserts that  every reasonably well-behaved function concentrates very sharply around its expectation value if we choose a random vector uniformly from a (real- or complex-valued) unit sphere $\mathbb{S}^{D-1}$ in $D \gg 1$ dimensions. Note that Haar-random state  $\ket{\psi} = U|\psi_0 \rangle$ with $U \sim \muU$ meet this sampling requirement by definition. 
The well-behavedness of functions is measured by their Lipschitz constant. 
A function $f: \mathbb{S}^{D-1} \to \mathbb{R}$ is Lipschitz with respect to the $\ell_2$-norm in $\mathbb{C}^D$ with constant $L \geq 0$  if
\begin{equation*}
\left| f(\ket{\psi}) - f(|\phi \rangle) \right| \leq L \left\| \ket{\psi} - |\phi \rangle \right\|_2 
\quad \text{for all $\ket{\psi}, |\phi \rangle \in \mathbb{S}^{D-1}$}.
\end{equation*}
Here is a variant of Levi's Lemma (concentration of measure) that directly applies to pure quantum states in $D$ dimensions. It readily follows from identifying the complex unit sphere $\mathbb{S}^{D-1} \subset \mathbb{C}^D$ with a real-valued unit sphere in $2D$ dimensions isometric embedding, see, e.g.
\cite[proof of Proposition~29]{brandao_complexity_2021}.

\begin{theorem}[Levy's lemma for Haar-random pure states] \label{thm:levy}
Let $f: \mathbb{S}^{D-1} \to \mathbb{R}$ be a function from $D$-dimensional pure states to the real numbers that is Lipschitz with Lipschitz constant $L$. Then,
\begin{align*}
\Pr_{\ket{\psi} \sim \muS} \left[\left| f(\ket{\psi}) - \Ex_{\ket{\psi} \sim \muS}\left[ f(\ket{\psi})\right] \right| >\tau \right] \leq 2 \exp \left( - \frac{4D\tau^2}{9 \pi^3 L^2} \right) \quad \text{for any $\tau>0$.}
\end{align*}
\end{theorem}

In words, Levy's lemma suppresses the probability of a $\tau$-deviation from the expectation value. This bound diminishes exponentially in Hilbert space dimension $D=2^n$, i.e.\ doubly exponentially in qubit size $n$. In the following two sections we will use \cref{thm:levy} to obtain bounds on the maximally distinguishable fraction and the probability of being far from uniform. We do so by explicitly upper bounding the Lipschitz constants of the involved functions and computing the Haar expectation values.

\subsection{Maximally distinguishable fraction}
Let us start with $\Pr_{U \sim \muU}[|P_U[\phi]-\mathcal{U}[\phi]|>\tau]$. We will use the short-hand notation $x=(x_1,\ldots,x_n) \in \left\{0,1\right\}^n$ to enumerate all possible outcome strings of $n$ parallel computational basis measurements:
\begin{align*}
P_U \left[ \phi \right] =& \sum_{x \in \left\{0,1\right\}^n} \phi (x_0,\ldots,x_n) \left| \braket{x}{\psi} \right|^2
= \mel{ \psi}{\lr{ \sum_{x \in \left\{0,1\right\}^n} \phi (x) \ketbra{x}{x}}}{\psi}
= \langle \psi |\Phi |\psi \rangle,
\end{align*}
where we have introduced the diagonal matrix $\Phi = \sum_{x} \phi (x) |x \rangle \! \langle x| \in \mathbb{C}^{D \times D}$
whose spectral norm obeys $\| \Phi \|_\infty = \max_{x \in \left\{0,1\right\}^n}|\phi (x)| \leq 1$ regardless of the underlying function in question.
This is a very simple and highly structured quadratic form in $\ket{\psi}=U\ket{\psi_0}$. 
Its expectation value over all Haar-random states produces the uniform distribution $\mathcal{U}[\phi]$, in formulas
\begin{equation}
\Ex_{U \sim \muU} \left[ P_U [\phi] \right] = \sum_{x \in \left\{0,1\right\}^n} \frac{1}{2^n} \phi (x) = \mathcal{U}[\phi]\,.
\label{eq:single-function-haar-expectation}
\end{equation}
To see this, note that the uniform average over all pure states in $D$ dimensions produces the maximally mixed state, i.e.\ $\Ex_{\ket{\psi}\sim\muS} \left[ \ketbra{\psi}{\psi}\right] = \mb1/D$. Linearity of the expectation value then ensures
\begin{align*}
\Ex_{U \sim \muU} \left[ P_U[\phi]\right]=& 
\Ex_{\ket{\psi} \sim\muS} \left[ \langle \psi| \Phi| \psi \rangle \right]= 
\Tr \left( \Ex_{\ket{\psi} \sim\muS}\left[ |\psi \rangle \! \langle \psi|\right]\; \Phi \right) = \mathrm{tr} \left( \mb1/D \; \Phi \right) = \frac{1}{D}\tr \left( \Phi \right) \\
=& \sum_{x} \frac{1}{2^n}\phi (x) = \mathcal{U}[\phi],
\end{align*}
as claimed. We also provide an alternative derivation based on Gaussian integration in the Appendix (\Cref{thm:function-expectation-app}). We can now obtain an explicit bound on the maximally distinguishable fraction by  \cref{thm:levy}.

\begin{theorem} \label{thm:haar-function}
Consider $n$-qubit Haar-random unitaries $U \sim \muU$ ($D=2^n$) and fix a function $\phi: \left\{0,1\right\}^n \to \left[-1,1\right]$. Then,
\begin{equation*}
\Pr_{U \sim \muU}\left[ \left| P_U [\phi]-\mathcal{U}[\phi] \right| >\tau \right] \leq 2 \exp \left( - \frac{2^{n} \tau^2}{9 \pi^3}\right) \quad \text{for any $\tau >0$}.
\end{equation*}
\end{theorem}
In words: for each fixed function $\phi$, its evaluation $P_U[\phi]$ concentrates very sharply doubly-exponentially in qubit size $n$ around the uniform average. Hence, it is extremely unlikely to distinguish $P_U$ from $\mathcal{U}$ with only a single $\phi$. 

\begin{proof}
Rewrite $P_U [\phi] = \langle \psi| \Phi| \psi \rangle=: f_\Phi(|\psi \rangle)$ with $\ket{\psi} = U |\psi_0 \rangle$ (Haar random state) and diagonal matrix $\Phi$ that obeys $\|\Phi \|_\infty \leq 1$. This reformulation highlights that the quadratic form function $f_\Phi: \mathbb{S}^{D-1} \to \mathbb{R}$ is Lipschitz with constant $L \leq 2 \|\Phi \|_\infty \leq 2$, we refer to \Cref{lem:lipschitz1} in the appendix for a detailed statement and proof. 
Next, recall from \Cref{eq:single-function-haar-expectation} that $\Ex_{|\psi\rangle \sim\muS} \left[ f_\Phi (|\psi \rangle)\right] = \Ex_{U \sim \muU} \left[ P_U [\phi]\right] = \mathcal{U}[\phi]$ and apply \Cref{thm:levy} to deduce the claim.
\end{proof}

\subsection{Far from uniform probability}
Moving now to the quantity $\Pr_{U \sim \muU}[\tv(P_U, \mathcal{U})\geq \epsilon]$, we note that the expectation value required is a bit more involved by comparison. Let us start by reformulating the total variation distance between $P_U$ and $\mathcal{U}$ as
\begin{align*}
\tv(P_U,\mathcal{U}) = \frac{1}{2}\sum_{x \in \left\{0,1\right\}^n} \left| P_U (x) - \mathcal{U}(x) \right| = \frac{1}{2}\sum_{x \in \left\{0,1\right\}^n} \left| \left| \langle x \mid U |\psi_0 \rangle \right|^2 - \frac{1}{D} \right|&\\ =
\frac{1}{2D}\sum_{x \in \left\{0,1\right\}^n} \left| \left| D\langle x \mid \psi \rangle \right|^2 - 1 \right|&,
\end{align*}
where $D=2^n$ and $|\psi \rangle = U |\psi_0 \rangle$. Linearity of the expectation value and unitary invariance of the spherical Haar measure  then implies
\begin{align*}
\Ex_{U \sim \muU}\left[\tv(P_U,\mathcal{U}) \right] = \frac{1}{2D} \sum_{x \in \left\{0,1\right\}^n} \Ex_{|\psi \rangle \sim\muS} \left[ \left| D \left| \langle x| \psi \rangle \right|^2 -1\right| \right] &\\
= \frac{1}{2} \Ex_{|\psi \rangle \sim\muS} \left[ \left| D \left|\langle 0,\ldots,0| \psi \rangle \right|^2-1 \right|\right]&.
\end{align*}
The expression on the right hand side is not a polynomial in the overlap $\left| \langle 0,\ldots,0 | \psi \rangle \right|^2$ which prevents us from using Weingarten calculus to compute it. Another averaging technique, known as Gaussian integration, does the job, however. The key idea is to view the uniform expectation over pure states as a uniform integral over all points that are contained in the complex-valued unit sphere in $D$ dimensions. 
Up to a normalization factor (scaling), this integral can then be re-cast as an expectation value over the directional degrees of freedom in a $2^n$-dimensional complex-valued Gaussian random vector with independent entries $g_j+\mathrm{i} h_j$, $1 \leq j \leq 2^n$ and $g_j,h_j \overset{\textit{iid}}{\sim} \mathcal{N}(0,1)$. A detailed argument is provided in the appendix and yields
\begin{align}
\frac{1}{\mathrm{e}}- \frac{1}{2^{n/2+1}} \leq \Ex_{U \sim \muU}\left[\tv(P_U,\mathcal{U}) \right]
\leq \frac{1}{\mathrm{e}} + \frac{1}{2^{n/2+1}}, \label{eq:TV-haar}
\end{align}
where $\mathrm{e}$ denotes Euler's constant. Note that the approximation errors on the left and right decay exponentially in qubit size $n$.
This is the content of \Cref{thm:haar-TV-app} in the appendix and the proof uses a precise version of the approximate identity
\begin{align*}
\frac{1}{2}\Ex_{|\psi \rangle \sim\muS} &\left[ \left| D \ryantext{|\langle 0,\ldots,0| \psi \rangle|^2}
-1 \right| \right] \\
&\approx \frac{1}{4} \Ex_{g_j,h_j\overset{\textit{iid}}{\sim}\mathcal{N}(0,1)} \left[ 
\left|  \left| g_1 + \mathrm{i}h_1 \right|^2 -2 \right| \right]
= \frac{1}{4}\Ex_{g_1,h_1 \overset{\textit{iid}}{\sim}\mathcal{N}(0,1)} \left[ \left|  g_1^2 + h_1^2  -2 \right| \right] \\
&= \frac{1}{4}\iint_\infty^\infty \left| g_1^2 + h_1^2 -2 \right| \frac{\exp \left( - (g_1^2+h_1^2)/2\right)}{2\pi} \mathrm{d}g_1 \mathrm{d}g_2 = \frac{1}{\mathrm{e}}.
\end{align*}
The final equality follows from switching into polar coordinates and solving the resulting integral analytically -- this is why the technique is called Gaussian integration. 
The exponentially small offsets $\pm 1/2^{n/2+1}$ in Rel.~\eqref{eq:TV-haar} 
bound the approximation error that is incurred in the first step of this argument. As before, we can now obtain a bound on the probability of a Haar randomly drawn unitary giving rise to a distribution that is far from uniform by use of 
\Cref{thm:levy}.

\begin{theorem}
Consider $n$-qubit Haar-random unitaries $U \sim \muU$ ($D=2^n$). Then, the TV distance between $P_U$ and the uniform distribution $\mathcal{U}$ is guaranteed to obey
\begin{equation*}
\Pr_{U \sim \muU}\left[ \left|\tv(P_U,\mathcal{U})- \frac{1}{\mathrm{e}}\right| \geq \xi + \frac{1}{2^{n/2+1}} \right]
\leq 2 \exp \left( - \frac{2^{n+2} \xi^2}{9 \pi^3} \right) \quad \text{for any $\xi >0$}.
\end{equation*}
\end{theorem}

In words: this TV distance concentrates very sharply (doubly-exponentially in qubit size $n$) around the remarkable value $1/\mathrm{e} \geq 0.367$.
The exponentially small in $n$ additive correction term $1/2^{n/2+1}$ is a consequence of the slight mismatches in Rel.~\eqref{eq:TV-haar}. 
Note, however, that this does not qualitatively change the concentration statement. The mismatch is of the same order as the smallest $\xi$ for which the exponential tail bound still provides meaningful results: $2^{n+2} \xi^2 \gtrsim 1$ requires $\xi \gtrsim 1/2^{n/2+1}$.

\begin{proof}
The proof is conceptually very similar to the proof of \Cref{thm:haar-function}.
We first recast the expectation over Haar-random unitaries $U$ as an expectation value over Haar-random state  $\ket{\psi} = U |\psi_0 \rangle$. The function $\tv(P_U,\mathcal{U})$ in question becomes
$
g(\ket{\psi}) = ({2D})^{-1}\sum_{x \in \left\{0,1\right\}^n} \left| D \left| \langle x | \psi \rangle \right|^2-1\right| 
$
and can be shown to be Lipschitz with constant $L \leq 1$. Again, we refer to \Cref{lem:lipschitz2} in the appendix for a precise statement and proof. Next, we use Rel.~\eqref{eq:TV-haar} to infer
\begin{equation*}
\Pr \left[ \left|  \tv(P_U,\mathcal{U})- \frac{1}{\mathrm{e}}\right| \geq \xi + \frac{1}{2^{n/2+1}}\right]
\leq \Pr \left[ \left| \tv (P_U,\mathcal{U}) -  \Ex_{U \sim \muU}\left[\tv(P_U,\mathcal{U}) \right] \right| \geq \xi \right] 
\end{equation*}
and apply Levy's Lemma (\Cref{thm:levy}) with Lipschitz constant $L=1$ to the right hand side of this display.
\end{proof}

\section{Random quantum circuits of linear depth}\label{sec:maximal_distinguishable}
In this section, we bound the two key quantities, namely $\mathfrak{f}$ and the far from uniform probability, for random quantum circuits of linear depth. 
We will find that the strong convergence of these circuit ensembles to unitary $t$-designs suffices to show exponentially small upper bounds on both quantities.

Plugging these bounds into \Cref{lem:deterministic-average-complexity-2}, we obtain the following lower bound on the average case query complexity.
\begin{theorem}[Formal version of linear depth part of \Cref{infthm:main-result}]
\label{thm:linear-depth-query-complexity}
        Let $\tau>0$. Further, let \ryantext{the circuit depth be}
        $d\geq1.2\times10^{20}n$,
        let $\epsilon \leq 1/150-\tau$ 
        and let $n$ be large enough. Then, the average case query complexity $q$ of $\epsilon$-learning any $\beta$-fraction of brickwork random quantum circuit output distributions is lower bounded by
    \begin{equation}
        q +1\geq\left(\beta -  3200\times 2^{-n}\right) 2^{n-2} \tau^2\,.
    \end{equation}
\end{theorem}

\begin{remark}
Note that for any $\tau\geq2^{-n/4}$ and, say any $\epsilon\leq1/160$ by \Cref{thm:linear-depth-query-complexity} learning a fraction $\beta>3200\cdot2{-n}=O(2^{-n})$ requires at least $q=2^{\Omega(n)}$ many queries.

Moreover, in the practically inspired regime $1/\poly(n)\leq\tau\leq1/150-\epsilon$, we obtain $q=\Omega(2^n)$.
\end{remark}

\subsection{Maximally distinguishable fraction}\label{sec:frac-linear-depth}

We begin with bounding the maximally distinguishable fraction.

\begin{lemma}
\label{l:frac_linear_depth}
Let $\delta>0$, $n\geq2$ and $d\geq 3.2((2+\ln(2))n+\ln(n)+\ln(1/\delta))$. Then for all $\phi:\{0,1\}^n\rightarrow [-1,1]$ it holds
\begin{equation}
    \underset{U\sim\muC}{\Pr}\lrq{ \abs{P_U[\phi] - \mathcal{U}[\phi]}> \tau } \leq \frac{(2+\delta)}{2^n \tau^2}.
\end{equation}
\end{lemma}

\begin{proof}
First, we show the result for an exact unitary $2$-design.
Using the first moment from \Cref{eq:Clifford_first_mom}, we find that 
\begin{equation}
 	\Ex_{U\sim \muU} \lrq{P_U[\phi]}
 	= \sum_{x\in\{0,1\}^n}\lr{\Ex_{U\sim \muU} \lrq{ P_U(x) } \phi(x)} = \mc U[\phi].
\end{equation}
Thus, by Chebyshev's inequality, for any $\tau>0$,
\begin{equation}
    \Pr_{U\sim\muU}\lrq{\abs{P_U[\phi] - \mc U[\phi]}> \tau } \leq \frac{\Var\lrq{P_U[\phi]}}{\tau^{2}}.
\end{equation}
The variance is given by
\begin{align}
	\Var_{U\sim\muU}\lrq{P_U[\phi]}
	&=\Ex_{U\sim\muU} \lrq{P_U[\phi]^2} - \lr{\Ex_{U\sim\muU} \lrq{ P_U[\phi]}}^2\\[5pt]
	&= \sum_{x}\sum_{y}\phi(x)\phi(y) \lr{\Ex_{U\sim\muU} \lrq{ P_U(x) P_U(y) }-\frac{1}{2^{2n}}}\,.
 \label{eq:variance}
\end{align}
Inserting the second moment from 
\Cref{eq:Clifford_second_mom} and bounding $\phi(x)\phi(y)\leq 1$, we find
\begin{align}\label{eq:var-bound-2-design}
	\Var_{U\sim\muU}\lrq{P_U[\phi]}
	&= \sum_{x}\sum_{y}\phi(x)\phi(y) \lr{ \frac{1}{2^n(2^n+1)}[1 + \delta_{x,y}] -\frac{1}{2^{2n}}}
	\leq \frac{1}{2^{n-1}}=O(2^{-n})
\end{align}
which holds for any exact unitary 2-design.

Let us now turn back to random quantum circuits.
At depth $d\geq 3.2((2+\ln(2))n+\ln(n)+\ln(1/\delta))$, the measure $\muC$ forms an $\delta\cdot 2^{-n}$-approximate $2$-design~\cite{haferkamp2021improved}.
Notice that by Hölder's inequality and \Cref{eq:statedesign}, it holds
\begin{align}
\begin{split}
     \Ex_{U\sim\muC}& \lrq{P_U(x)P_U(y)} - \Ex_{U\sim\muU} \lrq{P_U(x)P_U(y)}\\ &\leq 
     \abs{\Tr\lrq{\ketbra{x}{x}\otimes \ketbra{y}{y} \lr{\Ex_{U\sim\muC}\lrq{U\ketbra{0^n}{0^n}U^{\dagger}}^{\otimes 2}-\Ex_{U\sim\muU}\lrq{U\ketbra{0^n}{0^n}U^{\dagger}}^{\otimes 2}}}}\\
    &\leq \norm{\Ex_{U\sim\muC}\lrq{U\ketbra{0^n}{0^n}U^{\dagger}}^{\otimes 2}-\Ex_{U\sim\muU}\lrq{U\ketbra{0^n}{0^n}U^{\dagger}}^{\otimes 2}}_1\\
    &\leq \frac{\delta}{2^{3n}}\,.
    \end{split}
\end{align}

Then, as in \Cref{eq:var-bound-2-design}, we arrive at 
\begin{align}
\begin{split}
\Var_{U\sim\muC}[P_U[\phi]]
&=\sum_x\sum_y\phi(x)\phi(y)\lr{\Ex_{U\sim\muC}[P_U(x)P_U(y)]-\frac{1}{2^{2n}}} \\
&\leq \sum_x\sum_y\phi(x)\phi(y)\lr{\Ex_{U\sim\muU}[P_U(x)P_U(y)]+\frac{\delta}{2^{3n}}-\frac{1}{2^{2n}}}\\
&\leq \frac{1}{2^{n-1}}+\frac{\delta}{2^n}=\frac{2+\delta}{2^n}\,,
\end{split}
\end{align}
which completes the proof.
\end{proof}

\subsection{Far from uniformity via unitary designs}

In this section, we will use higher moments to show a far from uniform property that holds with probability $1-\exp(-\Omega(n))$.
Notice, that third moments cannot suffice to prove such a statement as the Clifford group is a $3$-design but a constant fraction ($\approx 0.4$) of stabilizer states yield uniform output distributions as shown in \Cref{app:far-from-uniform-global-cliffords}. 

A standard technique for lower bounding expectation values of $\ell_1$-norms is Berger's inequality~\cite{berger1997fourth}
\begin{equation}
    \Ex\lrq{\abs{S}}\geq \frac{\lr{\Ex\lrq{S^2}}^{\frac{3}{2}}}{\lr{\Ex\lrq{ S^4}}^{\frac{1}{2}}},
\end{equation}
for a random variable $S$.
However, applying this to the expected total variation distance yields another constant lower bound on the expectation value, which is not sufficient to prove a bound with probability $1-o(1)$.
Instead, we will use that, very similarly, we have:
\begin{lemma}\label{lemma:berger_for_norms}
\begin{equation}
\norm{ f} _{1}\geq\frac{\norm{ f}_{2}^{3}}{\norm{ f} _{4}^{2}}
\end{equation}
 for functions $f:\left\{ 0,1\right\} ^{n}\to\mathbb{R}$.
 \end{lemma}
 \begin{proof}
 This follows from Hölder's inequality
 \begin{equation}
     \Vert f\Vert_2^2=\langle f^{a},f^b\rangle\leq \left(\sum_{x}f^{pa}(x)\right)^{\frac{1}{p}}\left(\sum_{x}f^{qb}(x)\right)^{\frac{1}{q}},
 \end{equation}
 for $a+b=2$ and $\frac{1}{p}+\frac{1}{q}=1$.
 Choosing $a=\frac43$, $b=\frac23$, $p=3$ and $q=\frac32$, yields the result.
 \end{proof}
 
 We will prove concentration inequalities both for the numerator and the denominator using eighth moments and then apply a union bound to show that both scale as desired with high probability.
 This will allow us to prove a qualitative version of the conjecture by Aaronson and Chen~\cite{aaronsonComplexityTheoreticFoundationsQuantum2017a} in the affirmative.
 
 \begin{theorem}\label{thm:ball_U}
 Let $\mu$ be a  $ \frac{1}{1000}2^{-10n}$-approximate unitary $8$-design and $n\geq 2$, then:
\begin{equation}
\Pr_{U\sim\mu}\lrq{\tv\left(P_{U},\mathcal{U}\right)\geq \frac{1}{150}}\geq 1- 3200\times 2^{-n}.
\end{equation}
\end{theorem}
 Applications of the $8$-design property are rare and we pose it as an open problem whether \Cref{thm:ball_U} can be further derandomized\ryantext{, i.e., whether a similarly strong bound can be proved based on lower order moments.}
 
We know that $3$-designs are not sufficient as the Clifford group is one.
Can the same scaling be shown using only the (approximate) $4$-design property?
To this end, we prove the following theorem, which does not yield exponential concentration, but a weaker trade-off between the probability and the total variation distance.
This property therefore still allows for meaningful average-case hardness statements with probability $1-o(1)$, which separates $4$-designs from the Clifford group.
Moreover, far better constants are known for the generation of approximate $4$-designs~\cite{haferkamp2021improved}.

 \begin{theorem}\label{thm:4-design}
 Let $\mu$ be a  $ \frac{1}{1000}2^{-10n}$-approximate unitary $4$-design and $n\geq 2$, then for any $c>0$ we have:
\begin{equation}
\Pr_{U\sim\mu}\lrq{\tv\left(P_{U},\mathcal{U}\right)\geq \frac{1}{20\sqrt{c+18}}}\geq 1- 100\times 2^{-n}-\frac{25}{c}.
\end{equation}
\end{theorem}

Before we prove \Cref{thm:ball_U} and \Cref{thm:4-design}, we state the following corollary of \Cref{thm:ball_U} which is due to the bounds from \cite{haferkamp2021improved}:

\begin{corollary}\label{cor:far-from-uniform-linear}
Denote by $\muC$ the distribution on $\U(2^n)$ obtained from brickwork random quantum circuits of depth $d$.
For $d\geq 1.2\times 10^{20}n$ and $n\geq 2$, we have
\begin{equation}
    \Pr_{U\sim \muC}\lrq{\tv\left(P_{U},\mathcal{U}\right)\geq \frac{1}{150}}\geq 1- 3200\times 2^{-n}.
\end{equation}
\end{corollary}
\begin{proof}[Proof of \Cref{thm:ball_U}]
Applying \Cref{lemma:berger_for_norms} to the function $f(x)=P_{U}(x)-\mathcal{U}(x)$ we get
\begin{equation}\label{eq:Berger_to_tv_distance}
\norm{ P_{U}-\mathcal{U}} _{1}\geq\frac{\norm{ P_{U}-\mathcal{U}}_{2}^{3}}{\norm{ P_{U}-\mathcal{U}}_{4}^{2}}.
\end{equation}
The proof strategy is to show concentration inequalities for the numerator and the denominator, independently.
Then, we apply the union bound to show that both events are realized simultaneously with high probability.

We apply Chebyshev's inequality to the \textit{collision probability} $Z:=\sum_xP_U(x)^2$ in order to estimate $\norm{P_U-\mc U}_2^2=Z-1/D$.
We will use the notation $D=2^n$.
In the following, we will make an error of size $10^{-3} D^{-10}$, compared to the Haar value when evaluating monomials $\Ex\lrq{P_U(x_1)^{\lambda_1}\ldots P_U(x_k)^{\lambda_k}}$ with $\sum_k \lambda_k\leq 8$.
 In the following calculations, we denote by $E_i\in \mathbb{R}$ with $i=1,2,3,4$ error terms with $|E_i|\leq 10^{-3} D^{-10}$.
 We have chosen this error such that it is negligible for any of the below calculations.
 The reader may ignore it, but needs to keep in mind that it mildly affects the constants.

Using \Cref{lemma:momentcalculation}, we compute the first and second moment of $Z$
by
\begin{align}
\Ex_{U\sim\mu}\lrq{Z} = \frac{2}{D+1}+DE_1
\end{align}
and
\begin{align}
\begin{split}
    \Ex_{U\sim\mu} \lrq{Z^2}&=\Ex_{U\sim\mu}\lrq{\sum_{x\neq y}P_U(x)^2P_U(y)^2+\sum_{x}P_U(x)^4}\\
    &= \frac{4(D-1)}{(D+1)(D+2)(D+3)}+\frac{24}{(D+1)(D+2)(D+3)}+D^2 E_2\\
    &=\frac{4}{(D+1)(D+2)}+\frac{8}{(D+1)(D+2)(D+3)}+D^2 E_2\\
    &\leq 4D^{-2}+8 D^{-3}\,.
    \end{split}
\end{align}

We readily compute the variance $\sigma^2$ of Z:
\begin{align}
\begin{split}
    \sigma^2&=\Ex_{U\sim\mu}\lrq{Z^{2}}-\left(\Ex_{U\sim\mu}\lrq{Z}\right)^2\\
    &\leq 4D^{-2}+8D^{-3}-4\frac{1}{(D+1)^2} + 2D |E_1|+D^2|E_1|^2\\
    &\leq 17\times D^{-3}\,,
    \end{split}
\end{align}
where we have used in the last inequality that $1-x^2<1$ for $x>0$ implies $\frac{1}{1+x}>1-x$ and hence
\begin{equation}
    \left(\frac{1}{D+1}\right)^2= \left(\frac{1}{1+D^{-1}}\right)^2 D^{-2}\geq \left(1-D^{-1}\right)^2 D^{-2}\geq (1-2\times D^{-1})D^{-2}\,,
\end{equation}
where again the last step is Bernoulli's inequality.
Plugging this into Chebyshev's inequality, we find 
\begin{align}
\begin{split}
    \label{eq:chebyshev_to_Z}
    \Pr_{U\sim\mu}\lrq{\abs{\norm{P_U-\mathcal{U}}_2^2-\left(\frac{D-1}{D+1}\right)D^{-1}- D E_1}\geq k}=
    \Pr_{U\sim\mu}\lrq{\abs{Z-\frac{2}{D+1}-D E_1}\geq k}\\[5pt]
    =\frac{\sigma^2}{k^2}\leq \frac{17D^{-3}}{k^2}
\end{split}
\end{align}
for the probability of $Z$ being outside an interval of radius $k$ centered at its mean
Choosing $k=\frac{1}{2}\frac{D-1}{D+1}\cdot D^{-1}-DE_1$, this implies
\begin{align}\label{eq:Zisnottoosmall}
\begin{split}
\Pr_{U\sim\mu}\lrq{\norm{P_U-\mathcal{U}}_2^2\leq \frac12 \frac{D-1}{D+1} D^{-1} }&\leq 17\times \left(\frac{D-1}{D+1}D^{-1}-DE_1\right)^{-2}\times D^{-3}\\
&\leq 18\times \left(\frac{D+1}{D-1}\right)^2D^{-1}\\
&\leq 18\times \left(\frac{2^1+1}{2^1-1}\right)^2D^{-1}\\
&\leq 50 D^{-1}\,.
\end{split}
\end{align}
Here we used in the second inequality that $\frac{1}{1-x}\leq 1+2x$ for $0\leq x\leq \frac12$, which, after multiplying it with $1-x$ is equivalent to $0\leq x(1-2x)$.
This bound will be sufficient to bound the numerator of \Cref{eq:Berger_to_tv_distance}.

Next, we will find a concentration inequality for the denominator of \Cref{eq:Berger_to_tv_distance} by essentially the same strategy.
We find 
\begin{align}
\begin{split}\label{eq:upperboundonX}
\norm{P_U-\mathcal{U}}_4^4&=\sum_x\lr{P_U(x)-D^{-1}}^4\\
&=
\sum_x\lr{P_U(x)^4-4P_U(x)^3 D^{-1}+6P_U(x)^2 D^{-2}-4P_U(x) D^{-3}+D^{-4}}\\
&\leq  \underbrace{\sum_x P_U(x)^4}_{=:X} +6D^{-2}\underbrace{\sum_x P_U(x)^2}_{=Z}\,,
\end{split}
\end{align}
where we have used $\sum_x \lr{D^{-4} -4P_U(x)D^{-3}}\leq0$.

We will prove probability inequalities for the two terms in \Cref{eq:upperboundonX} independently.
In fact, the second term can be handled similarly to \Cref{eq:Zisnottoosmall}.
We apply \Cref{eq:chebyshev_to_Z} with $k=\frac{1}{D+1}-DE_1$ to obtain
\begin{equation}\label{eq:Zisnottoobig}
    \Pr_{U\sim\mu}\lrq{Z\geq \frac{3}{D+1}\times D^{-1}}\leq 18 (D+1)^2D^{-3}\leq 30 D^{-1}.
\end{equation}

For the first term in \Cref{eq:upperboundonX} we use \Cref{lemma:momentcalculation} to compute the first
\begin{align}
    \begin{split}
        \Ex_{U\sim\mu}[X]=\sum_x\Ex_{U\sim\mu}\lrq{P_U(x)^4}=\frac{4!}{D\cdots(D+3)}+DE_3&\\
        =\frac{24}{(D+1)(D+2)(D+3)}+DE_3&\,,
    \end{split}
\end{align}
 and second moments
\begin{align}
\begin{split}
\Ex_{U\sim\mu} \lrq{X^2}&=\sum_{x}\Ex_{U\sim\mu} \lrq{P_U(x)^8}+\sum_{x\neq y}\Ex_{U\sim\mu} \lrq{P_U(x)^4P_U(y)^4}\\
&=\frac{8!}{(D+1)\cdots (D+7)}+\frac{ 24^2(D-1)}{(D+1)\cdots (D+7)}+D^2 E_4\\
&=\frac{8!-2\times 24^2}{(D+1)\cdots (D+7)}+\frac{ 24^2}{(D+2)\cdots (D+7)}+D^2 E_4\,.
\end{split}
\end{align}

For the variance this implies
\begin{align}
\begin{split}
\sigma_X^2&\leq \frac{8!}{(D+1)\cdots (D+7)}+\frac{ 24^2}{(D+2)\cdots (D+7)}+D^2E_4\\
&\quad-\frac{24^2}{(D+1)^2(D+2)^2(D+3)^2}+2\frac{24^2}{(D+1)^2(D+2)^2(D+3)^2} D |E_3|+D^2E_3^2\\[10pt]
&\leq 41000 D^{-7}+24^2\underbrace{\lr{\frac{1}{(D+2)\cdots (D+7)}-\frac{1}{(D+1)^2(D+2)^2(D+3)^2}}}_{<0}\\
&\leq 41000 D^{-7}.
\end{split}
\end{align}
Via Chebyshev's inequality, we obtain 
\begin{equation}
\Pr_{U\sim\mu}\lrq{
\abs{X-\frac{24}{(D+1)(D+2)(D+3)}+DE_3}\geq k 
}
\leq \frac{41000D^{-7}}{k^2}.
\end{equation}
Choosing $k=\frac{12}{(D+1)(D+2)(D+3)}-DE_3$, this implies
\begin{equation}\label{eq:Yisnottoobig}
\Pr_{U\sim\mu}\lrq{X\geq \frac{36}{(D+1)(D+2)(D+3)}}\leq \frac{41001}{12^2}D^{-7}(D+1)^2(D+2)^2(D+3)^2\leq 3100 D^{-1}\,,
\end{equation}
for $n\geq 2$.

We can now put these bounds together via a union bound applied twice:
For any $\eps_1,\eps_2>0$ such that $\frac{\eps_1}{\eps_2}\geq \eps$, we find
\begin{align}
\begin{split}\label{eq:unionboundapplied}
 \Pr_{U\sim\mu}\lrq{\norm{ P_{U}-\mathcal{U}}_{1}\leq \eps}&\leq \Pr_{U\sim\mu}\lrq{\frac{\norm{ P_{U}-\mathcal{U}}_{2}^{3}}{\norm{ P_{U}-\mathcal{U}}_{4}^{2}}\leq \eps}\\
 &\leq \Pr_{U\sim\mu}\lrq{\left(\norm{ P_{U}-\mathcal{U}}_{2}^{3}\leq \eps_1\right)\vee \left(\norm{ P_{U}-\mathcal{U}}_{4}^{2}
 \geq
 \eps_2\right) }\\
 &\leq \Pr_{U\sim\mu}\lrq{\norm{ P_{U}-\mathcal{U}}_{2}^{3}\leq \eps_1} +\Pr_{U\sim\mu}\lrq{\norm{ P_{U}-\mathcal{U}}_{4}^{2}\geq \eps_2}\\
  &= \Pr_{U\sim\mu}\lrq{\norm{ P_{U}-\mathcal{U}}_{2}^{2}\leq \eps_1^{\frac{2}{3}}} +\Pr_{U\sim\mu}\lrq{\norm{ P_{U}-\mathcal{U}}_{4}^{4}\geq \eps_2^{2}}\\
   &\leq \Pr_{U\sim\mu}\lrq{\norm{ P_{U}-\mathcal{U}}_{2}^{2}\leq \eps_1^{\frac{2}{3}}} +\Pr_{U\sim\mu}\lrq{X+6D^2(Z+2^{-n})\geq \eps_2^{2}}\\
  &\leq \Pr_{U\sim\mu}\lrq{\norm{ P_{U}-\mathcal{U}}_{2}^{2}\leq \eps_1^{\frac{2}{3}}} +\Pr_{U\sim\mu}\lrq{X\geq a_1} +\Pr_{U\sim\mu}\lrq{6D^{-2}Z\geq a_2}\,,
 \end{split}
\end{align}
with $a_1+a_2=\epsilon_2^2$.
These probabilities are bounded in \Cref{eq:Zisnottoosmall,eq:Zisnottoobig,eq:Yisnottoobig}.
Choosing $\eps_1^{\frac{2}{3}}= \frac{1}{2}\frac{D-1}{D+1} D^{-1}\geq 10^{-\frac23}D^{-1}$ (for $n\geq 2$), $a_1=36\times D^{-3}$ and $a_2=18\times D^{-3}$ implies $\eps_2^2=54\times D^{-3}$.
Plugging this into \Cref{eq:unionboundapplied} yields \Cref{thm:ball_U}, where $\tv\left(P_{U},\mathcal{U}\right)=\frac12 ||P_U-\mathcal{U}||_1$.
\end{proof}

\begin{proof}[Proof of \Cref{thm:4-design}]
The proof of \Cref{thm:4-design} follows analogously to the proof of \Cref{thm:ball_U}.
However, instead of proving concentration of the random variable $X=\sum_{x} P_U(x)^4$ to bound $\Pr[X\geq a_1]$ as in \Cref{eq:unionboundapplied}, we instead apply Markov's inequality,
to get
\begin{equation}
\Pr_{U\sim\mu}[X\geq a_1]\leq \frac{\Ex_{U\sim\mu}[X]}{a_1}\leq \frac{24}{(D+1)(D+2)(D+3)}\cdot\frac{1}{a_1}+DE_3\leq \frac{25 D^{-3}}{a_1}.
\end{equation}
Therefore, the result follows by choosing $a_1=c D^{-3}$ for $c>0$.
\end{proof}

\section{Random quantum circuits of sub-linear depth}\label{sec:far_from_uniform}

In this section, we bound the two key quantities, namely $\mathfrak{f}$ and the far from uniform probability, for random quantum circuits in the regime of sub-linear depth. This regime of circuit depth includes the shortest circuits for which we can still show super-polynomial query complexity lower bounds and hence hardness of learning. Note that in this regime, the random circuit ensembles that we consider do not yet form even a unitary 2-design requiring different techniques to obtain these bounds.
Plugging these bounds into \Cref{lem:deterministic-average-complexity-2}, we obtain the following lower bound on the average case query complexity.

\begin{theorem}[Formal version of sub-linear depth part of \Cref{infthm:main-result}]
\label{thm:query-complexity-sublinear-depth}
    Let $c=1/\log(5/4)$.
    Further, let $ c\log n\leq d\leq c(n+\log n)$, $\tau>0$ and $\epsilon \leq 1/4-\tau$. Then, for sufficiently large $n$, the average case query complexity $q$ of $\epsilon$-learning any $\beta$-fraction of brickwork random quantum circuits is lower bounded by
    \begin{equation}
        q+1 \geq \frac{ \lr{ \beta - 3/4 - \epsilon -\tau} \tau^2}{3n} \cdot \lr{\frac45}^d\,,
    \end{equation}
    Moreover, if $d>c(n+\log n)$ then it holds
    \begin{align}
        q+1\geq\lr{\beta-3/4-\epsilon -\tau}2^{n-2}\tau^2 \,.
    \end{align}
\end{theorem}

\begin{remark}
Note that \Cref{thm:query-complexity-sublinear-depth} implies that in the practically relevant regime of $\tau=1/\poly(n)$, learning circuits of depth $\omega(\log(n))$ to some constant precision requires a super-polynomial number of queries.
In particular, already for any constant fraction $\beta$ slightly greater than $3/4$ there exists no efficient statistical query algorithm for any accuracy $\tau=\Omega(1/\poly(n))$.
\end{remark}

\subsection{Maximally distinguishable fraction via restricted depth moments}\label{sec:frac-superlog-depth}

We begin with bounding the maximally distinguishable fraction.

\begin{lemma}
\label{l:frac_superlog_depth}
For all $d\geq \frac{\log n}{\log 5/4}$ and for all $\phi:\{0,1\}^n\rightarrow [-1,1]$ it holds
\begin{align}
    &\Pr_{U\sim\muC}\lrq{ \abs{P_U[\phi] - \mathcal{U}[\phi]}> \tau } \leq \frac{1}{\tau^2} \left[ n \left(\frac{4}{5}\right)^d\left( 1+\frac{1}{2^n} \right)+\frac{1}{2^n} \right]\,.\label{eq:sublinear}
\end{align}
\end{lemma}

\begin{remark}
Note that \Cref{eq:sublinear} can be simplified as follows, which lead to the two cases in \Cref{thm:query-complexity-sublinear-depth}:
\begin{align}
    \Pr_{U\sim\muC}\lrq{ \abs{P_U[\phi] - \mathcal{U}[\phi]}> \tau } \leq
    \begin{cases}
        \frac{3n}{\tau^2}\cdot\lr{\frac45}^d\;,\quad&\text{for}\;d\leq\frac{ n+\log(n)}{\log(5/4)}\\
        \frac{3}{2^n\tau^2}\;,\quad&\text{for}\;d>\frac{ n+\log(n)}{\log(5/4)}\,.
    \end{cases}
\end{align}
\end{remark}

\begin{proofof}[\Cref{l:frac_superlog_depth}]
The proof strategy is essentially identical to the first part of the proof of \Cref{l:frac_linear_depth} bounding the same quantity for linear depth circuits. However, one replaces the moments over the Haar measure with the moments of restricted depth models obtained in \Cref{l:brickwork_moments}. In particular, the second moment in \Cref{eq:variance} gets replaced by the one given in \Cref{eq:brickwork_second_mom}. The moments in \Cref{l:brickwork_moments} hold for restricted depth circuits in 1D. They are obtained via a mapping to a statistical mechanics model \cite{hunterjones19} also used in \cite{barak2021spoofing} to bound a similar quantity, for more details see \Cref{app:moments}.
\end{proofof}

\subsection{Far from uniformity for constant-depth circuits}

A direct approach to bound $\Pr\lrq{\tv(P_U,\mc U)\geq\eps}$ is to lower bound the expectation 
$\Ex\lrq{\tv(P_U,\mc U)}$. Then, using Markov's inequality, a far-from-uniform-bound follows immediately as made explicit by the following lemma:

\begin{lemma}
 \label{lem:far_from_uniform_via_markov}For any random variable
$0\leq Y\leq1$ and any $0<\epsilon<1$ we have
\begin{equation}
\Pr\lr{Y\geq\epsilon}\geq\frac{\Ex\lrq{Y}-\epsilon}{1-\epsilon}\,.
\end{equation}
\end{lemma}

In \cite{aaronsonComplexityTheoreticFoundationsQuantum2017a}, Aaronson and Chen show the following lower bound on $\Ex_{U\sim\muC}\lrq{\tv(P_U,\mc U)}$.

\begin{lemma}[Section 3.5,  \cite{aaronsonComplexityTheoreticFoundationsQuantum2017a}]
For any $n\geq2, d\geq1$, it holds that
\begin{equation}
    \Ex_{U\sim\muC}\lrq{\tv(P_U,\mc U)}\geq\frac{1}{4}\,.
\end{equation}
\end{lemma}

Curiously, their proof only takes into account the randomness in drawing the very last two-qubit gate which is why the bound holds already at depth $d=1$.
By \Cref{lem:far_from_uniform_via_markov}, we immediately find:

\begin{corollary}\label{cor:const_far_from_uniform}
   For any $n\geq2, d\geq1,\epsilon\in[0,1/4]$, it holds that
   \begin{equation}
        \Pr_{U\sim\muC}\lrq{{\tv(P_U,\mc U)\geq\eps}}
      \geq\frac{\frac{1}{4}-\epsilon}{1-\epsilon} \geq \frac{1}{4} - \epsilon \,.
   \end{equation}
\end{corollary}

\section*{Acknowledgements}
We thank Yihui Quek, Dominik Hangleiter, Jean-Pierre Seifert, Scott Aaronson, and Lijie Chen for many helpful discussions and insights. 
J.~H.~acknowledges funding from the Harvard Quantum Initiative. The work of R.~K.~is supported by the SFB BeyondC (Grant No. F7107-
N38), the Project QuantumReady (FFG 896217) and the BMWK (ProvideQ).
The Berlin team thanks the BMWK (PlanQK, EniQmA), the BMBF (Hybrid), and the Munich Quantum Valley (K-8). This work has also been funded by the Deutsche Forschungsgemeinschaft (DFG) under Germany's Excellence Strategy – The Berlin Mathematics Research Center MATH+ (EXC-2046/1, project ID: 390685689) as well as CRC 183 (B1).

%%%%%%
%
%  APPENDIX
%
%%%%%%

\appendix

\section{Omitted proofs for Haar random unitaries}

In this appendix section, we provide the technical details for studying the Haar random case.
We first introduce Gaussian integration -- a powerful way to compute expectation values of functions over uniformly random pure states Haar random states. 
This technique is complementary to the more widely used Weingarten formalism. The two approaches have different strengths and weaknesses. One core advantage of Gaussian integration is that it can be readily applied to functions that are not well approximated by homogeneous polynomials. 
The expected TV distance between a Haar random outcome distribution and the uniform distribution is one such function which features prominently in this work.

Subsequently, we will also provide tight bounds on the Lipschitz constants of these functions. This is the only missing ingredient to show exponentially strong concentration around the previously computed expectation values by means of Levy's lemma.

We emphasize that this appendix section covers the extreme case of global Haar-random unitaries. Such evolutions scramble information across all subsystems and we therefore present our results directly in terms of total Hilbert space dimension $D$, an abstract computational basis $\left\{|1 \rangle, \ldots, |D \rangle \right\}$ and global (pure) states $|\psi \rangle = U|\psi_0 \rangle \in \mathbb{C}^D$, where $|\psi_0 \rangle$ is a designated (simple) starting state, e.g., a product state. For $n$-qubit systems, this would amount to setting $D=2^n$ and equating the $d$-th basis state $|d \rangle$ with the $n$-bit string $\llcorner d \lrcorner = x_1(d) \ldots x_n(d) \in \left\{0,1\right\}^n$ that corresponds to the binary representation of $d \in \mathbb{N}$.

\subsection{Haar random state averages via Gaussian integration}

Gaussian integration is a standard technique in mathematical signal processing and data science that allows us to recast an expectation value over the unit sphere as an expectation value over standard Gaussian random vectors. The latter has the distinct advantage that individual vector components become statistically independent from each other and follow the very simple normal distribution. 
These features are very useful for computing non-polynomial functions that only depend on comparatively few vector entries. 
We will showcase this technique based on two exemplary expectation values that feature prominently in this work: 
\begin{itemize}
\item[(i)] the function value $P_U \left[\phi\right]=\sum_{d=1}^D \phi (d) \left| \langle d| U |\psi_0 \rangle \right|^2$ averaged uniformly over all unitaries $U$, see \Cref{thm:function-expectation-app} below and 
\item[(ii)] the expected TV distance between the outcome distribution of a Haar random pure state $|\psi \rangle=U|\psi_0 \rangle$ and the uniform distribution, see \Cref{thm:haar-TV-app} below. 
\end{itemize}
At the heart of both computations is the following powerful meta-theorem, known as Gaussian integration.

\begin{theorem}[Gaussian integration] \label{thm:Gaussian-integration}
Let $f: \mathbb{C}^D \to \mathbb{C}$ be a homogeneous function with even degree $2k \in 2\mathbb{N}$, i.e., $f( a \mathbf{x})= |a|^{2k} f(\mathbf{x})$ for $\mathbf{x} \in \mathbb{C}^D$ and $a \in \mathbb{C}$. Then,
we can rewrite the Haar expectation value as 
\begin{equation}
\Ex_{|\psi \rangle \sim\muS} \left[ f (|\psi \rangle) \right] 
= 
\frac{1}{k! 2^k} \binom{D+k-1}{k}^{-1}
\Ex_{g_i,h_i\overset{\textit{iid}}{\sim} \mathcal{N}(0,1)} \left[ f \left( g_1 + \mathrm{i} h_1,\ldots,g_D + \mathrm{i}h_D\right) \right],
\label{eq:Gaussian-integration}
\end{equation}
where $g_i,h_i \overset{\textit{iid}}{\sim} \mathcal{N}(0,1)$ denote independent standard Gaussian random variables ($\mu=0$, $\sigma^2=1$).

\end{theorem}
In words: the left hand side features the Haar average over all pure states (complex unit vectors), while the right hand side features an expectation value over $2D$ statistically independent standard Gaussian random variables $g_j,h_j\sim \mathcal{N}(0,1)$ with probability density function $\exp \left( - x^2/2 \right)/\sqrt{2\pi}$. This reformulation can be extremely helpful in concrete calculations, because the individual vector components on the right hand side are all statistically independent.

\begin{proof}
Let us start by considering the complex-valued standard Gaussian vector $\mathbf{g}+\mathrm{i}\mathbf{h} \in \mathbb{C}^D$ with $\mathbf{g}=(g_1,\ldots, g_D), \mathbf{h} = (h_1,\ldots,h_D) \in \mathbb{R}^D$ and $g_j,h_j \overset{\textit{iid}}{\sim}\mathcal{N}(0,1)$. Switching to polar coordinates allows us to rewrite this vector as
\begin{align}
\mathbf{g} + \mathrm{i} \mathbf{h} = r (\mathbf{g},\mathbf{h}) | \widehat{\mathbf{g}+\mathrm{i} \mathbf{h}} \rangle\quad 
\intertext{with direction $|\widehat{\mathbf{g}+\mathrm{i} \mathbf{h}}  \rangle \in \mathbb{S}^{D-1}$ and radius} 
r(\mathbf{g},\mathbf{h})^2 = \sum_{j=1}^D \left( g_j^2 + h_j^2 \right)
\end{align}
We can now use homogeneity of the function $f: \mathbb{C}^D \to \mathbb{C}$ to rewrite the Gaussian expectation value on the right hand side of \Cref{eq:Gaussian-integration} as
\begin{align}
\Ex_{\mathbf{g},\mathbf{h} \sim \mathcal{N}(0,\mathbb{I})} \left[ f (\mathbf{g}+\mathrm{i} \mathbf{h}) \right] 
= \Ex_{\mathbf{g},\mathbf{h}} \left[ f \left( r(\mathbf{g},\mathbf{h}) |\widehat{\mathbf{g}+\mathrm{i} \mathbf{h}} \rangle \right) \right] 
= \Ex_{\mathbf{g},\mathbf{h}} \left[ r (\mathbf{g},\mathbf{h})^{2k} f \left( |\widehat{\mathbf{g}+\mathrm{i} \mathbf{h}}\rangle \right) \right],
\end{align}
where we have succinctly accumulated all Gaussian random variables into two vectors $\mathbf{g}$ and $\mathbf{h}$.
Now, something interesting happens. The particular form of the standard Gaussian probability density ensures that radius ($r(\mathbf{g},\mathbf{h})$) and direction ($|\widehat{\mathbf{g}+\mathrm{i} \mathbf{h}}\rangle$) 
can be decomposed into two statistically independent random variables / vectors. The square $r^2$ of the former follows a $\chi^2_{2D}$-distribution with $2D$ degrees of freedom while the latter 
must be a unit vector that is sampled uniformly from the complex unit sphere $\mathbb{S}^{D-1}$. The latter is a consequence of the fact that the distribution of standard Gaussian vectors $\mathbf{g}+\mathrm{i} \mathbf{h}$ is invariant under unitary transformations. Statistical independence, on the other hand, can be deduced from transforming the probability density function (pdf) of a Gaussian random vector into generalized spherical coordinates. Under such a transformation, the pdf decomposes into a product of a radial part (the radius) and an angular part.
We can now use these insights to decompose the original expectation value into a product of two expectation values:
\begin{align}
\begin{split}
    \Ex_{\mathbf{g},\mathbf{h}\overset{\textit{iid}}{\sim}\mathcal{N}(0,\mathbb{I})} \left[ r (\mathbf{g},\mathbf{h})^{2k} f \left( |\widehat{\mathbf{g}+\mathrm{i} \mathbf{h}}\rangle \right) \right]
    =& \Ex_{r^2 \sim \mathcal{\chi}^2_{2D}, \,\widehat{\mathbf{g}+\mathrm{i} \mathbf{h}} \sim\muS} \left[ r^{2k} f \left( |\widehat{\mathbf{g}+\mathrm{i}\mathbf{h}} \rangle \right) \right] \\
    =& \Ex_{r^2 \sim \mathcal{\chi}^2_{2D}} \left[ r^{2k} \right] \Ex_{ \widehat{\mathbf{g}+\mathrm{i} \mathbf{h}} \sim\muS} \left[ f \left( \widehat{\mathbf{g}+\mathrm{i}\mathbf{h}} \rangle \right) \right].
\end{split}
\end{align}
The second expression describes a uniform integral of $f$ over the complex unit sphere in $D$ dimensions. This is exactly the Haar expectation value on the left hand side of \Cref{eq:Gaussian-integration}. The other expression is the $k$-th moment of a $\chi^2$ random variable with $2D$ degrees of freedom. These are well-known and amount to
\begin{equation*}
\Ex_{r^2 \sim \chi^2_{2D}} \left[ r^{2k} \right] = \frac{\Gamma (2D+k/2)}{\Gamma (k/2)}=(2D) (2D+2) \cdots (2D+2(k-1)) = k! 2^k \binom{D+k-1}{k}.
\end{equation*}
Putting everything together, allows us to conclude
\begin{align}
\begin{split}
\Ex_{\mathbf{g},\mathbf{h}\overset{\textit{iid}}{\sim}\mathcal{N}(0,\mathbb{I})} \left[ f \left( \mathbf{g}+\mathrm{i} \mathbf{h}\right)\right] =&
\Ex_{g_i,h_i \overset{\textit{iid}}{\sim} \mathcal{N}(0,1)} \left[ f \left( g_1+\mathrm{i}h_1,\ldots,g_D + \mathrm{i} h_D \right) \right] \\
=&\Ex_{r^2 \sim \chi^2_{2D}} \left[ r^{2k} \right] \Ex_{|\psi \rangle \sim\muS} \left[ f( |\psi \rangle) \right] \\
=& k! 2^k \binom{D+k-1}{k} \Ex_{|\psi \rangle \sim\muS} \left[ f( |\psi \rangle) \right].
\end{split}
\end{align}
The claim in \Cref{eq:Gaussian-integration} is an immediate reformulation of this equality.
\end{proof}

We now have the essential tool at hand to compute the Haar expectation values that matter for this work.

\begin{theorem}[Haar average of bounded function expectation values] \label{thm:function-expectation-app}
Let $\phi: \left\{1,\ldots,D\right\} \rightarrow \left[-1,1\right]$ be a function and define $P_{U} \left[ \phi \right] = \sum_{d=1}^{D} \phi (d) \left| \langle d| U |\psi_0 \rangle \right|^2$.
Then, the expectation value of $P_U[\phi]$ over Haar random unitaries becomes
\begin{equation}
\Ex_{U \sim \muU} \left[ P_U [\phi] \right] = \Ex_{|\psi \rangle \sim\muS} \left[ \sum_{d=1}^D \phi (d) \left| \langle d | \psi \rangle \right|^2 \right] = \sum_{d=1}^D \frac{1}{D} \phi (d) = \mathcal{U} \left[ \phi \right].
\end{equation}
\end{theorem}

This is a standard result that readily follows from Haar integration via Weingarten calculus and the representation theory of the unitary group. Let us now show how to achieve the same result with Gaussian integration (\Cref{thm:Gaussian-integration}). 

\begin{proof}
Let us start by using linearity to rewrite the desired expectation value as
\begin{align}
\Ex_{U \sim \muU} \left[ P_U \left[ \phi \right] \right] = \sum_{d=1}^D \phi (d) \Ex_{|\psi \rangle \sim\muS} \left[ \left| \langle d| \psi \rangle \right|^2 \right].
\end{align}
The claim then follows from the following equality:
\begin{equation}
\Ex_{|\psi \rangle \sim\muS} \left[ \left| \langle d| \psi \rangle \right|^2 \right] = \frac{1}{D} \quad \text{for all $d=1,\ldots,D$.}
\label{eq:app-function-expectation}
\end{equation}
which we now prove for any basis vector $d$.
Once $d$ is fixed, we can interpret this as the expectation value of the function $f_d (|\psi \rangle) = \left| \langle d| \psi \rangle \right|^2$ which selects the $d$'th vector entry (amplitude) and outputs its magnitude squared. Every such function is homogeneous with even degree $2$ ($k=1$) and we can use \Cref{thm:Gaussian-integration} to rewrite the expectation value as 
\begin{align*}
\Ex_{|\psi \rangle \sim\muS} \left[ \left| \langle d| \psi \rangle \right|^2 \right]
=& \Ex_{|\psi \rangle \sim\muS} \left[ f_d (|\psi \rangle) \right]\\
=& \frac{1}{1! 2^1} \binom{D}{1}^{-1} \Ex_{g_i,h_i \overset{\textit{iid}}{\sim} \mathcal{N}(0,1)} \left[ f_d (g_1 + \mathrm{i} h_1,\ldots,g_D + \mathrm{i} h_D)\right] \\
=& \frac{1}{2D} \Ex_{g_i,h_i \overset{\textit{iid}}{\sim} \mathcal{N}(0,1)} \left[ \left| g_d + \mathrm{i} h_d \right|^2 \right] 
= \frac{1}{2D} \left( \Ex_{g_d \sim \mathcal{N}(0,1)} \left[ g_d^2 \right] + \Ex_{h_d \sim \mathcal{N}(0,1)} \left[ h_d^2 \right] \right) \\
=& \frac{1}{2D} \left(1 + 1 \right) = \frac{1}{D}.
\end{align*}
Here, we have used statistical independence (we can forget all Gaussian random variables which do not feature in the expression), as well as the fact that Gaussian standard variables obey $\Ex\left[ g_d^2 \right] = \Ex \left[ h_d^2 \right] =1$ (unit variance). 
\end{proof}

The next expectation value is more intricate by comparison. There, Weingarten calculus would not work, because the function involved is not a well-behaved polynomial. Gaussian integration, however, can handle such non-polynomial expectation values and yields the following elegant display.

\begin{theorem}[Haar expectation of the TV distance] \label{thm:haar-TV-app}
Let $P_U(d)=\abs{\mel{d}{U}{\psi_0}}^2$ be the output distribution of a $D$ level quantum system in the computational basis and $\mc U$ the $D$ dimensional uniform distribution.
The expectation value of the total variation distance $\tv(P_U,\mc U)$ over Haar random unitaries $U$ obeys
\begin{equation*}
\frac{1}{\mathrm{e}} - \frac{1}{2\sqrt{D}} \leq \Ex_{U \sim \muU} \left[ \tv\left( P_U,\mathcal{U} \right) \right] \leq \frac{1}{\mathrm{e}} + \frac{1}{2\sqrt{D}}\,.
\end{equation*}
The approximation error is controlled by $1/(2\sqrt{D})$ which diminishes exponentially for $n$-qubit systems ($D=2^n$).
\end{theorem}

\begin{proof}
Let us once more start by using linearity of the expectation value to rewrite the desired expression as
\begin{align}
\Ex_{U \sim \muU} \left[ \tv \left( P_U, \mathcal{U} \right) \right] 
= \frac{1}{2} \sum_{d=1}^D \Ex_{|\psi \rangle \sim\muS} \left[ \frac{1}{2}\sum_{d=1}^D \left| \left| \langle d| \psi \rangle \right|^2 - \frac{1}{D} \right| \right]&\notag\\
= \frac{1}{2D} \sum_{d=1}^D \Ex_{|\psi \rangle \sim\muS} \left[ \left| D \left| \langle d| \psi \rangle \right|^2 -1 \right| \right]&.
\end{align}
Next, note that unitary invariance of $|\psi \rangle\sim\muS$ ensures that each of the summands on the right hand side must yield the same expectation value. This allows us to simplify further and obtain
\begin{align*}
\Ex_{U \sim \muU} \left[ \tv \left( P_U, \mathcal{U} \right) \right] 
= \frac{1}{2D} \sum_{d=1}^D \Ex_{|\psi \rangle \sim\muS} \left[ \left| D \left| \langle d| \psi \rangle \right|^2 -1 \right| \right]
= \frac{1}{2} \Ex_{|\psi \rangle \sim\muS} \left[ \left| D \left| \langle 1| \psi \rangle \right|^2 -1 \right| \right].
\end{align*}
Note that this is not a polynomial function, but we can still use Gaussian integration to accurately bound this expression.
To this end, 
we first use $\langle \psi| \psi \rangle =1$ to rewrite the expression of interest as the uniform average over a homogeneous function with even degree $2$ ($k=1$):
\begin{align*}
f \left( |\psi \rangle \right) = \frac{1}{2}\left|D \left| \langle 1| \psi \rangle \right|^2 - \langle \psi |\psi \rangle \right|^2.
\end{align*}
We can now use \Cref{thm:Gaussian-integration} to conclude
\begin{align}
\Ex_{U \sim \muU} \left[ \tv \left( P_U, \mathcal{U} \right) \right] 
=&  \Ex_{|\psi \rangle \sim\muS} \left[ \frac{1}{2} \left| D \left| \langle 1 |\psi \rangle \right|^2 - \langle \psi| \psi \rangle \right| \right]
= \Ex_{|\psi \rangle \sim\muS}\left[ f(|\psi \rangle ) \right] \nonumber \\
=& \frac{1}{1! 2^1} \binom{D}{1}^{-1} \Ex_{g_i,h_i \overset{\textit{iid}}{\sim}\mathcal{N}(0,1)} \left[ f \left( g_1 + \mathrm{i} h_1, \ldots, g_d + \mathrm{i} h_d \right) \right] \nonumber\\
=& \frac{1}{2D} \Ex_{g_j,h_j \overset{\textit{iid}}{\sim} \mathcal{N}(0,1)} \left[ \frac{1}{2}\left| \left|D \langle 1 | \mathbf{g}+\mathrm{i} \mathbf{h} \rangle \right|^2 - \langle \mathbf{g} + \mathrm{i} \mathbf{h} | \mathbf{g} + \mathrm{i} \mathbf{h} \rangle \right| \right] \nonumber \\
=& \frac{1}{4D} \Ex_{g_j,h_j \overset{\textit{iid}}{\sim} \mathcal{N}(0,1)} \left[ \left| D \left( g_1^2 + h_1^2 \right) - \sum_{j=1}^D \left( g_j^2 + h_j^2 \right) \right| \right] \nonumber \\
=& \frac{1}{2} \Ex_{g_j,h_j \overset{\textit{iid}}{\sim} \mathcal{N}(0,1)} \left[ \left|\frac{1}{2} \left( g_1^2 + h_1^2 \right) - 1 + 1 - \frac{1}{2D} \sum_{j=1}^D \left( g_j^2 + h_j^2 \right) \right| \right].
\label{eq:TV-aux1}
\end{align}
It seems possible to compute this expectation value directly by rewriting each expectation value as an integral weighted by the Gaussian probability density function $\exp \left( - g_j^2/2\right)$, but doing so would incur a total of $2D$ nested integrations. Here, we instead simplify the derivation considerably by providing reasonably tight upper and lower bounds. We have already suggestively decomposed the expression within the absolute value into two terms that are easier to control individually:
\begin{align*}
M =& \frac{1}{2} \Ex_{g_1,h_1 \overset{\textit{iid}}{\sim} \mathcal{N}(0,1)} \left[ \left| \frac{1}{2}\left(g_1^2+h_1^2 \right) -1 \right| \right] & \text{(asymptotic mean value)}, \\
\Delta =& \frac{1}{2} \Ex_{g_j,h_j \overset{\textit{iid}}{\sim} \mathcal{N}(0,1)} \left[ \left| 1- \frac{1}{2D} \sum_{j=1}^D \left( g_j^2 + h_j^2 \right) \right| \right] & \text{(approximation error)}.
\end{align*}
Applying the triangle inequality to \Cref{eq:TV-aux1} readily allows us to infer
\begin{equation}
M - \Delta \leq \Ex_{U \sim \muU} \left[ \tv \left( P_U, \mathcal{U} \right) \right]  \leq M+ \Delta \label{eq:TV-approximation}
\end{equation}
and the two remaining parameters can now be computed independently. We defer the actual calculations to the end of this subsection and only state the results here:
\begin{align*}
M =& 1/\mathrm{e}& \text{(see \Cref{lem:M-app} below)}, \\
\Delta \leq &1/  \left(2 \sqrt{D}\right)& \text{(see \Cref{lem:Delta-app} below)}.
\end{align*}
Inserting these numerical values into \Cref{eq:TV-approximation} yields the claim.
\end{proof}

Let us now supply the technical calculations that are essential 
for 
completing the proof of \Cref{thm:haar-TV-app}.

\begin{lemma} \label{lem:Delta-app}
Let $g_1,\ldots,g_d,h_1,\ldots,h_d$ be $2D$ independent standard Gaussian random variables. 
Then,
\begin{equation*}
\Delta = \frac{1}{2} \Ex_{g_i,h_i \overset{\textit{iid}}{\sim} \mathcal{N}(0,1)} \left[ \left| 1 - \frac{1}{2D} \sum_{j=1}^D \left( g_j^2 + h_j^2 \right) \right| \right] \leq 
\frac{1}{2\sqrt{D}}.
\end{equation*}
\end{lemma}

\begin{proof}
There is no conceptual difference between the $g_j$ and $h_j$ random variables. So, we may replace them with $2D$ independent standard Gaussian random variables $\tilde{g}_1,\ldots,\tilde{g}_{2D} \overset{\textit{iid}}{\sim} \mathcal{N}(0,1)$. This reformulation yields
\begin{equation}
\Delta = \frac{1}{4D} \Ex_{\tilde{g}_i \overset{\textit{iid}}{\sim} \mathcal{N}(0,1)}  \left[ \left| 2D - \sum_{j=1}^{2D} \tilde{g}_j^2 \right| \right]
\leq \frac{1}{4D} \Ex_{\tilde{g}_i \overset{\textit{iid}}{\sim} \mathcal{N}(0,1)} \left[ \left( 2D- \sum_{j=1}^{2D} \tilde{g}_j^2 \right)^2 \right]^{1/2}, \label{eq:Delta-derivation1}
\end{equation}
where the last inequality is Jensen's. The remaining expectation value is the variance of a $\chi^2$ random variable with $2D$ degrees of freedom. Standard textbooks tell us that this variance is equal to $2 (2D)=4D$. We do, however, believe that it is instructive to compute this variance directly, because it showcases an important subroutine when computing Haar integrals via Gaussian integration. 
Note that the random variables involved obey 
\begin{align*}
    \Ex_{\tilde{g}_j,\tilde{g}_k \overset{\textit{iid}}{\sim} \mathcal{N}(0,1)} \left[ \tilde{g}_j^2 \tilde{g}_k^2 \right] = 
    \begin{cases}
    \Ex_{\tilde{g}_j \sim \mathcal{N}(0,1)} \left[ \tilde{g}_j^2 \right] \Ex_{\tilde{g}_k \sim \mathcal{N}(0,1)} \left[ \tilde{g}_k^2 \right] = 1 \,,\;\;&\text{whenever $j \neq k$ and}\\[10pt]
    \Ex \left[ \tilde{g}_j^4 \right] =3\,,\quad&\text{else if $j=k$.}
    \end{cases}
\end{align*}
Combining them yields 
$\Ex_{\tilde{g}_j,\tilde{g}_k \overset{\textit{iid}}{\sim} \mathcal{N}(0,1)} \left[ \tilde{g}_j^2 \tilde{g}_k^2 \right] = 1 + 2 \delta_{j,k}$.

If we also recall $\Ex_{\tilde{g}_j \sim \mathcal{N}(0,1)} \left[ \tilde{g}_j^2 \right]=1$, we can readily conclude
\begin{align*}
\Ex_{\tilde{g}_i \overset{\textit{iid}}{\sim} \mathcal{N}(0,1)} \left[ \left( 2D - \sum_{j=1}^{2D} \tilde{g}_j^2 \right)^2 \right] 
=& 4D^2 - 4D \sum_{j=1}^{2D} \Ex_{\tilde{g}_j \sim \mathcal{N}(0,1)} \left[ \tilde{g}_j^2 \right] + \sum_{j=1}^{2D}\sum_{k=1}^{2D} \Ex_{\tilde{g}_j,\tilde{g}_k \overset{\textit{iid}}{\sim} \mathcal{N}(0,1)} \left[ \tilde{g}_j^2 \tilde{g}_k^2 \right] \\
=& 4D^2 - 4D \sum_{j=1}^{2D} 1 + \sum_{j=1}^{2D} \sum_{k=1}^{2D} \left(1+2 \delta_{j,k} \right) \\
=& 4D^2 - 8D^2 + 4D^2  +4D = 4D.
\end{align*}
Inserting this variance expression into \Cref{eq:Delta-derivation1} yields the claim.
\end{proof}

\begin{lemma} \label{lem:M-app}
Let $g,h$ be two independent standard Gaussian random variables. Then
\begin{align*}
M = \frac{1}{2} \Ex_{g,h \overset{\textit{iid}}{\sim} \mathcal{N}(0,1)} \left[ \left| \frac{1}{2} \left( g^2 + h^2 \right) -1 \right| \right] = \frac{1}{\mathrm{e}}.
\end{align*}
\end{lemma}

\begin{proof}
This is the one location in our derivation, where we really utilize the power of Gaussian integration. We start by rewriting the expectation value as an integral over two independent standard Gaussian random variables with (standard Gaussian) probability density functions $\exp \left( - g^2/2\right)/\sqrt{2\pi}$ and $\exp \left( - h^2/2\right)/\sqrt{2\pi}$, respectively:
\begin{align*}
M =&  \frac{1}{2} \Ex_{g,h \overset{\textit{iid}}{\sim} \mathcal{N}(0,1)} \left[ \left| \frac{1}{2} \left( g^2 + h^2 \right) -1 \right| \right] 
= \frac{1}{4} \iint_{-\infty}^\infty \left| g^2 + h^2 -2 \right| \frac{\exp \left( - (g^2+h^2)/2\right)}{2\pi} \mathrm{d} g \mathrm{d} h.
\end{align*}
Next, we view $(g,h) \in \mathbb{R}^2$ as a $2$-dimensional vector and switch into polar coordinates: $(r \cos (\phi),$ $r \sin (\phi))$. Note that $g^2+h^2 =r^2$ and there is no angle dependence in the integral. Accordingly the volume element changes from $\mathrm{d}g \mathrm{d}h$ to $r \mathrm{d}r \mathrm{d} \phi$ and we obtain
\begin{align*}
M = \frac{1}{8 \pi}  \int_0^{2\pi} \int_0^\infty\left| r^2 -2 \right| \mathrm{e}^{-r^2/2} r \mathrm{d}r \mathrm{d} \phi
= \frac{1}{4} \int_0^{\infty} \left| r^2-2 \right|r \mathrm{e}^{-r^2/2} r \mathrm{d} r,
\end{align*}
where we have carried out the integral over the angle $\phi$ which cancels the $1/(2\pi)$-term in front of the expression. 
Next, we note that the sign of the absolute value changes as we change the integration range. For $r \in \left[0,\sqrt{2}\right]$, we have $\left|r^2-2\right| = (2-r^2)$ while $\left|r^2-2\right| = (r^2-2)$ for $r \in \left[\sqrt{2},\infty\right)$. This implies
\begin{align}
M =& \frac{1}{4} \int_0^{\infty} \left| r^2-2 \right|r \mathrm{e}^{-r^2/2} r \mathrm{d} r \notag\\
=& \frac{1}{4} \left( \int_0^{\sqrt{2}} \left( 2-r^2 \right) r \mathrm{e}^{-r^2/2} \mathrm{d} r + \int_{\sqrt{2}}^\infty \left(r^2-2\right) r \mathrm{e}^{-r^2/2} \mathrm{d}r \right) \nonumber \\
=& \frac{1}{4} \left( 2 \int_0^{\sqrt{2}} r \mathrm{e}^{-r^2/2} \mathrm{d}r - \int_0^{\sqrt{2}} r^3 \mathrm{e}^{-r^2/2} \mathrm{d}r + \int_{\sqrt{2}}^{\infty} r^3 \mathrm{e}^{-r^2/2} - 2 \int_{\sqrt{2}}^{\infty} r \mathrm{e}^{-r^2/2} \mathrm{d}r \right).
\label{eq:Gaussian-integrals-app}
\end{align}
These four remaining integrals can be determined from the following well-known Gaussian integration formulas for $a \leq b$:
\begin{align*}
\int_a^b r \mathrm{e}^{r^2/2} = \mathrm{e}^{-a^2/2} - \mathrm{e}^{-b^2/2} \quad \text{and} \quad \int_a^b r^3 \mathrm{e}^{-r^2/2} \mathrm{d}r = (a^2+2) \mathrm{e}^{-a^2/2} - (b^2+2) \mathrm{e}^{-b^2/2}.
\end{align*}
The limit $b \to \infty$ produces a vanishing contributions, because $\lim_{b \to \infty} \mathrm{e}^{-b^2/2}=0$ and $\lim_{b \to \infty} (b^2+2) \mathrm{e}^{-b^2/2}=0$
Inserting these values into \Cref{eq:Gaussian-integrals-app} yields
\begin{align*}
M =& \frac{1}{4} \left( 2 \left( \mathrm{e}^{-0} - \mathrm{e}^{-1} \right) - \left( (2+0)\mathrm{e}^{-0} -(2+2) \mathrm{e}^{-1} \right) + \left( (2+2) \mathrm{e}^{-1} - 0 \right) - 2 \left( \mathrm{e}^{-1} - 0 \right) \right) \\
=& \frac{1}{4} \left( 2 - 2/\mathrm{e} - 2 + 4/\mathrm{e} + 4/\mathrm{e} -2/\mathrm{e} \right) = \frac{1}{\mathrm{e}}.
\end{align*}
\end{proof}

\subsection{Lipschitz constants for function evaluations and TV distances}

In this appendix section, we derive Lipschitz constants for the functions whose expectation value value we computed in the previous subsection. 
We will see that these Lipschitz constants are small ($L=2$ and $L=1$, respectively) which is the only requirement we need to invoke Levy's lemma to show exponential concentration around these expectation values.

\begin{lemma} \label{lem:lipschitz1}
To start fix $\phi: \left\{1,\ldots,D\right\} \to \left[-1,1\right]$ and reinterpret  $P_U [ \phi] =\sum_{d=1}^D \phi(d) |\langle d| U |\psi_0 \rangle|^2$ as a function 
in the pure state $|\psi \rangle = U |\psi_0 \rangle$, namely $P_\phi \left( |\psi \rangle \right) = \sum_{d=1}^D \phi (d) \left| \langle d| \psi \rangle \right|^2$. This function has Lipschitz constant $L=2$, namely
\begin{align*}
\left| P_\phi (|\psi \rangle) - P_\phi (|\chi \rangle) \right| \leq 2 \left\| |\psi \rangle - |\chi\rangle \right\|_{\ell_2} \quad \text{for all pure states $|\psi \rangle, |\chi\rangle \in \mathbb{C}^D$.}
\end{align*}
\end{lemma}

\begin{proof}
Let us start by rewriting $P_\phi (|\psi \rangle)$ as a linear function in the (pure) density matrix $|\psi \rangle \! \langle \psi|$:
\begin{align}
P_\phi (|\psi \rangle) = \sum_{d=1}^D \phi (d) \left| \langle d| \psi \rangle \right|^2 = \langle \psi| \left(\sum_{d=1}^D \phi (d) |d \rangle \! \langle d| \right) |\psi \rangle
= \mathrm{tr} \left( \Phi \; |\psi \rangle \! \langle \psi| \right),
\end{align}
where we have introduced the diagonal $D \times D$ matrix $\Phi = \sum_{d=1}^D \phi (d) |d \rangle \! \langle d|$. Note that this matrix has operator norm $\| \Phi \|_\infty = \max_{1 \leq d \leq D} \left| \phi (d) \right| \leq 1$, because the function values $\phi (d)$ are confined to $\left[-1,1\right]$ by assumption. 
The matrix Hoelder inequality then implies
\begin{align*}
\left| P_\phi (|\psi \rangle) - P_\phi (|\chi\rangle) \right|
= \left| \mathrm{tr} \left( \Phi \; |\psi \rangle \! \langle \psi| \right) - \mathrm{tr} \left( \Phi \; |\chi \rangle \! \langle \chi| \right) \right|
= \left| \mathrm{tr} \left( \Phi \; \left( |\psi \rangle \! \langle \psi| -|\chi \rangle \! \langle \chi| \right)  \right)\right|&\\
\leq \|\Phi \|_\infty \left| |\psi \rangle \! \langle \psi| - |\chi \rangle \! \langle \chi| \right\|_1&,
\end{align*}
where $\| \cdot \|_1$ denotes the trace norm. Since $\| \Phi \|_\infty \leq 1$, the claim -- Lipschitz constant $L=2$ --  then follows from the following relation between trace distance of pure states and Euclidean distance of the state vectors involved:
\begin{equation}
\frac{1}{2} \left\| |\psi \rangle \! \langle \psi| - |\chi \rangle \! \langle \chi| \right\|_1 \leq  \left\| |\psi \rangle - |\chi \rangle \right\|_{2} \quad \text{for pure states $|\psi \rangle, |\chi \rangle \in \mathbb{C}^D$}. \label{eq:nuclear-to-euclidean}
\end{equation}
Let us now derive this useful relation. One way is to use the Fuchs-van de Graaf inequalities (which are tight for pure states) to relate the trace distance to a pure state fidelity:
\begin{align}
\frac{1}{2}\left\| |\psi \rangle \! \langle \psi| - |\chi \rangle \! \langle \chi| \right\|_1 =  \sqrt{1 - F \left( |\psi \rangle, |\chi \rangle\right)} =  \sqrt{1- \left|\langle \psi| \chi \rangle \right|^2}.
\end{align}
Finally, we can use $\left| \langle \psi| \chi \rangle \right| \leq 1$, as well as $\langle \psi |\psi \rangle = \langle \chi |\chi \rangle =1$ and $2\left|\langle \psi | \chi \rangle \right| \geq 2 \mathrm{Re} \left( \langle \psi |\chi \rangle \right) = \langle \psi |\chi \rangle + \langle \chi |\psi \rangle$ to obtain
\begin{align}
\sqrt{1- \left| \langle \psi |\chi \rangle \right|^2} =& \sqrt{1 + \left| \langle \psi| \chi \rangle \right|} \sqrt{1 - \left| \langle \psi |\chi \rangle \right|}
\leq \sqrt{1+1} \sqrt{1- \mathrm{Re} \left( \langle \psi| \chi \rangle \right)}\notag\\
=& \sqrt{1 - \langle \psi| \chi \rangle - \langle \chi |\psi \rangle +1} 
= \sqrt{\langle \psi| \psi \rangle - \langle \psi| \chi \rangle - \langle \chi| \psi \rangle + \langle \chi |\chi \rangle}\notag\\
=& \sqrt{ \left( \langle \psi| - \langle \chi|\right) \left( |\psi \rangle - |\chi \rangle \right)} =
\norm{ |\psi \rangle - |\chi \rangle }_{2}.
\end{align}
\end{proof}

\begin{lemma} \label{lem:lipschitz2}
Let $P_U(d)=\abs{\mel{d}{U}{\psi_0}}^2$ be the output distribution of a $D$ level quantum system in the computational basis and let $\mc U$ be the $D$ dimensional uniform distribution.
The total variation distance between both distributions defines a function in the pure state $\ket\psi = U \ket{\psi_0}$, namely $f_{\mathrm{TV}}(\ket\psi) =\frac12\sum_d\abs{\abs{\braket{d}{\psi}}^2-1/D}$. This function has Lipschitz constant $L=1$, i.e.,
\begin{equation*}
\left| f_{\mathrm{TV}} \left( |\psi \rangle \right) - f_{\mathrm{TV}} \left( |\chi \rangle \right) \right| \leq  \left\| |\psi \rangle - |\chi \rangle \right\|_{2} \quad \text{for all pure states $|\psi \rangle, |\chi \rangle \in  \mathbb{C}^D$.}
\end{equation*}
\end{lemma}

\begin{proof}
Let us start by rewriting the absolute value of the difference of different function values as
\begin{align}
&\left| f_{\mathrm{TV}} \left( |\psi \rangle \right) - f_{\mathrm{TV}} \left( |\chi \rangle \right) \right| = \notag\\
&\quad= \frac{1}{2} \left| \sum_{d=1}^D \left( \left| \left|\langle d| \psi \rangle \right|^2 - 1/D \right| - \left| \left| \langle d|\chi \rangle \right|^2 -1/D\right| \right) \right| \notag\\
&\quad= \frac{1}{2} \left| \sum_{d=1}^D \left( \left| \left| \left(\langle d| \chi \rangle \right|^2 -1/D\right) + \left( \left| \langle d| \psi \rangle \right|^2 - \left| \langle d| \chi \rangle \right|^2 \right) \right| - \left| \left| \langle d|\chi \rangle \right|^2 -1/D\right|^2 \right) \right| \notag\\
&\quad\leq \frac{1}{2}  \sum_{d=1}^D \left| \left| \langle d| \psi \rangle \right|^2 - \left| \langle d|\chi \rangle \right|^2 \right|,
\end{align}
where the last inequality follows from applying the triangle inequality to each summand in order to break up the two contributions in the first absolute value of the second line. The first contribution then cancels with the final term and we obtain the advertised display. We can now rewrite this new expression as
\begin{equation}
\frac{1}{2} \sum_{d=1}^D \left| \left| \langle d| \psi \rangle \right|^2 - \left| \langle d |\chi \rangle \right|^2 \right|
= \frac{1}{2} \sum_{d=1}^D \left| \langle d| \left( |\psi \rangle \! \langle \psi| - |\chi \rangle \! \langle \chi| \right) |d \rangle \right|,
\end{equation}
which accumulates the sum of the absolute values of the diagonal entries of the (pure) state difference $\left(|\psi \rangle \! \langle \psi| - |\chi \rangle \! \langle \chi|\right)$.
This sum of absolute diagonal entries is always smaller than the trace norm of the matrix in question\footnote{This relation is well known in matrix analysis and follows, for instance, from Helstrom's theorem.}. This relation implies 
\begin{equation}
\left| f_{\mathrm{TV}} \left( |\psi \rangle \right) - f_{\mathrm{TV}} \left( |\chi \rangle \right) \right|\leq  \frac{1}{2} \sum_{d=1}^D \left| \langle d| \left( |\psi \rangle \! \langle \psi| - |\phi \rangle \! \langle \phi| \right) |d \rangle \right| \leq \frac{1}{2} \left\| |\psi \rangle \! \langle \psi| - |\phi \rangle \! \langle \phi| \right\|_1,
\end{equation}
and the claim -- Lipschitz constant $L=1$ --  now follows from reusing \Cref{eq:nuclear-to-euclidean} to convert this trace norm distance into a Euclidean distance of the state vectors involved.
\end{proof}

\section{Unitary designs}\label{appendix:unitarydesigns}
In this appendix we provide context for approximate unitary designs; a key tool for the results in this work. Moreover, we discuss recent bounds on the generation of designs by random quantum circuits.
Recall the definition of the moment operator:
\begin{equation}
    \Phi^{(t)}(\nu)(A):=\int U^{\otimes t} A (U^{\dagger})^{\otimes t}\mathrm{d}\nu(U).
\end{equation}
\begin{definition}[$\varepsilon$-Approximate Design]
A probability distribution $\nu$ over $\U(D)$ is an $\varepsilon$-approximate unitary design if 
\begin{equation}
 \norm{\Phi^{(t)}(\nu)-\Phi^{(t)}(\muU)}_{\Diamond}\leq \frac{\varepsilon}{D^{t}},
\end{equation}
where $\norm{\bullet}_{\Diamond}$ denotes the diamond norm, or channel distinguishability, defined as the stabilized $1\to 1$ norm.
\end{definition}

In this work we will only be concerned with averages over states, that is the case 
$A=(\ketbra{\psi})^{\otimes t}$.
In this case we have the standard formula (see e.g.~\cite{harrow2013church}).
\begin{equation}
    \Phi^{(t)}(\muU)(A)=\int U^{\otimes t}
    \lr{\ketbra{\psi}}^{\otimes t}
    (U^{\dagger})^{\otimes t}\mathrm{d}\muU(U)
    =\frac{P_{\mathrm{sym},t}}{\binom{D+t-1}{t}},
\end{equation}
where $P_{\mathrm{sym},t}$ we denote the projector onto the symmetric subspace $S^t(\mb{C}^D)$.
With the above definition of an approximate unitary design, we obtain that for $\nu$ an $\varepsilon$-approximate unitary design, we have 
\begin{equation}\label{eq:statedesign}
    \norm{\Phi^{(t)}(\nu)\lr{(\ketbra{\psi})^{\otimes t}}-\frac{P_{\mathrm{sym},t}}{\binom{D+t-1}{t}}}_1\leq \frac{\varepsilon}{D^t},
\end{equation}
where $\norm{\bullet}_1$ denotes the Schatten $1$-norm, or trace norm.

The key result for the following is that random quantum circuits are in fact approximate unitary $t$-designs in 
polynomial depth~\cite{brandao2016local, harrow2009random, haferkamp2022random}. 
These bounds come with large explicit constants.
For small values of $t=2,4$, we even have good explicit constants.
We present the bound from Haferkamp~\cite{haferkamp2022random}:
\begin{theorem}
For $n\geq \left\lceil2\log_2(4t)+1.5\sqrt{\log_2(4t)}\,\right\rceil$, random quantum circuits in a brickwork architecture are $\varepsilon$-approximate unitary $t$-designs in depth
\begin{equation}
    T\geq C\ln^5(t)t^{4+3\frac{1}{\sqrt{\log_2(t)}}}(2nt+\log_2(1/\varepsilon)),
\end{equation}
where $C$ can be taken to be $10^{13}$.
\end{theorem}
Note that the large constants are likely an artefact of the proof technique based on the martingale technique~\cite{nachtergaele1996spectral} in \cite{brandao2016local,haferkamp2022random}, which focus on the scaling in $t$.

Using instead finite-size criteria~\cite{knabe1988energy} combined with numerics, one can greatly improve these constants for $t\leq 5$. Compare {\cite[Table~I]{haferkamp2021improved}}.
It is likely that we could obtain comparable constants for $t=8$ as well.
Unfortunately, this seems to require numerics for daunting system sizes.

\section{Moment calculations}
\label{app:moments}

\subsection{Haar moments}\label{app:haar-moments}

To begin we give explicit formulas for Haar random moments.
We will make use of the following standard formula repeatedly:
\begin{align*}
\begin{split}
    \Ex_{\psi\sim \muS}\lrq{\abs{\braket{\psi}{\phi}}^{2t}} &= \Ex_{\psi\sim \muS}\lrq{\Tr\lrq{\lr{\ketbra{\psi}}^{\otimes t} \lr{\ketbra{\phi}}^{\otimes t}}}\\
    &=\Tr\lrq{\Ex_{\psi\sim \muS}\lrq{\ketbra{\psi}}^{\otimes t} \lr{\ketbra{\phi}}^{\otimes t}}\\
        &=\Tr\lrq{\frac{P_{\mathrm{sym},t,D}}{\binom{D+t-1}{t}}
        \lr{\ketbra{\phi}}^{\otimes t}}\label{eq.t-th_moment}\\
        &=\binom{D+t-1}{t}^{-1}\,.
    \end{split}
\end{align*}

In fact, we will need a more general formula for the proof of Theorem~\ref{thm:ball_U}, which we state as the 
following lemma.

\begin{lemma}\label{lemma:momentcalculation}
Let $|i_1\rangle,\ldots,|i_k\rangle$ with $i_1,\ldots,D\in \{1,\ldots, D\}$ be mutually orthogonal state vectors and $\lambda=(\lambda_1,\dots, \lambda_k)$ a partition of $t$ for $t\leq D$.
Then, we find the formula
\begin{equation}
     \Ex_{\ket{\psi}\sim\muS}\lrq{\prod_{l=1}^k |\langle\psi|i_l\rangle|^{2\lambda_l}}=\frac{\prod_{l=1}^k \lambda_l !}{D\cdots (D+t-1)}.
\end{equation}
\end{lemma}
\begin{proof}
The proof follows directly from the following calculation:
\begin{align}
\begin{split}
    \Ex_{\ket{\psi}\sim\muS}\prod_{l=1}^k |\langle\psi|i_l\rangle|^{2\lambda_l}
    &
    =\Ex_{\psi\sim \muS}\lrq{\Tr\left[(|\psi\rangle\langle\psi|)^{\otimes t}\bigotimes_{l=1}^k (|i_l\rangle\langle i_l|)^{\otimes \lambda_l} \right]}\\
    &=\binom{D+t-1}{t}^{-1} \Tr\left[P_{\mathrm{sym},t,D} \bigotimes_{l=1}^k (|i_l\rangle\langle i_l|)^{\otimes \lambda_l}\right]\\
    &= \frac{1}{D\cdots (D+t-1)}\sum_{\pi\in S_t} \Tr\left[r(\pi) \bigotimes_{l=1}^k (|i_l\rangle\langle i_l|)^{\otimes \lambda_l}\right],
    \end{split}
\end{align}
where we used the notation $r$ for the representation of $S_t$ that, for each permutation $\pi\in S_t$, permutes the $t$ tensor factors according to $\pi$:
\begin{equation}
    r(\pi)|i_1\rangle\otimes\cdots\otimes|i_t\rangle:=|i_{\pi^{-1}(1)}\rangle\otimes \cdots\otimes|i_{\pi^{-1}(t)}\rangle.
\end{equation}
Moreover, we used the formula $P_{\mathrm{sym},t,D}=\frac{1}{t!} \sum_{\pi} r(\pi)$.
Notice that 
\begin{equation}
    \Tr\left[r(\pi) \bigotimes_{l=1}^k (|i_l\rangle\langle i_l|)^{\otimes \lambda_l}\right]=\begin{cases}
    1\qquad\text{if}~~\pi\in S_{\lambda_1}\times \ldots\times S_{\lambda_{k}}\\
    0\qquad \text{else}.
    \end{cases}
\end{equation}
Hence, we find
\begin{align}
\begin{split}
    \Ex_{\ket{\psi}\sim\muS}\prod_{l=1}^k |\langle\psi|i_l\rangle|^{2\lambda_l}&= \frac{1}{D\cdots (D+t-1)} |S_{\lambda_1}|\cdots |S_{\lambda_k}|\\
    &=\frac{\prod_{l=1}^k \lambda_l !}{D\cdots (D+t-1)}.
\end{split}
\end{align}
\end{proof}

In the special case $D=2^n$ we thus obtain the explicit formulas for the first and second moment:
\begin{align}
    	\Ex_{U\sim\muU} \left[ P_U(x) \right] 
	&=\frac{1}{2^n} \label{eq:Clifford_first_mom},\\
	\Ex_{U\sim\muU} \left[ P_U(x) P_U(y) \right] 
	& =\frac{1}{2^n(2^n+1)}[1 + \delta_{x,y}]\,.
	\label{eq:Clifford_second_mom}
\end{align}

\subsection{Restricted depth moments} \label{app:restricted-depth-moments}

Next, we state bounds on the first two moments over brickwork random quantum circuits of depth $d$:

\begin{lemma}[Moments over circuits of restricted depth -- adapted from \cite{barak2021spoofing}] \label{l:brickwork_moments}
For $\muC$ the measure over brickwork random quantum circuits on $n$ qubits of depth $d$, it holds
\begin{align}
    \Ex_{U\sim \muC}	\left[ P_U(x) \right] 
	&= \frac{1}{2^n} \label{eq:brickwork_first_mom_2} \,,\\
 \Ex_{U\sim \muC} \left[ P_U(x) P_U(y) \right] 
	&\leq 
	\frac{1}{2^{2n}}
	(1+\delta_{x,y})\left[1+n\left(\frac{4}{5}\right)^{d}\right]\,.\label{eq:brickwork_second_mom}
\end{align}
 where the bound in \Cref{eq:brickwork_second_mom} holds for $d\geq \frac{\log n}{\log 5/4}$.
\end{lemma}

\begin{proof}
We note that $\muC$ is an exact 1-design at any depth $d$\footnote{In fact, already a single layer of randomly drawn unitary gates forms an exact 1-design. This is because this layer contains as a subgroup the Pauli group which is known to form an exact 1-design. It follows from the invariance of the Haar measure under left multiplication that random unitary circuits form an exact 1-design also for $d\geq1$.}. 
Hence, the first moment is the same as in \Cref{eq:Clifford_first_mom} i.e.
\begin{equation}
 \Ex_{U\sim\muC} \left[ P_U(x) \right] 
	= \Ex_{U\sim\muU} \left[ P_U(x) \right] = \frac{1}{2^n}.
\end{equation}
To obtain the second moment given in \Cref{eq:brickwork_second_mom}, we adapt and modify a calculation presented in Section 6.3 of \cite{barak2021spoofing}.  Specifically, using a mapping to a statistical mechanics model, the second moment with respect to the random circuit, $\Ex_{\muC} \left[ P_U(x) P_U(y) \right] $, can be expressed as a partition function. The value of this partition function can then be bounded by counting domain walls. In Section 6.3 of Ref. \cite{barak2021spoofing}, this technique was already used to obtain an upper bound on $\Ex_{\muC} \left[ P_U(x)^2 \right]$, for random circuits of depth $d\geq \frac{\log n}{\log 5/4}$. More specifically, Ref. \cite{barak2021spoofing} has obtained the upper bound

\begin{equation}
	\Ex_{U\sim\muC} \left[ P_U(x)^2 \right]\leq \left(1+\left(\frac{4}{5}\right)^{d}\right)^{n/2} \Ex_{U\sim\muU}\left[ P_U(x)^2 \right],
\end{equation}
which is given in terms of the Haar expectation value $\Ex_{\muU} \left[ P_U(x)^2 \right]$, 
and indeed converges to this Haar value in the infinite circuit depth-limit $d\rightarrow \infty$. 
A similar analysis allows us to obtain the following bound on the expectation value of the cross terms $P_U(x) P_U(y)$,
\begin{equation}
	\Ex_{U\sim\muC} \left[ P_U(x) P_U(y) \right] \leq \left(1+\left(\frac{4}{5}\right)^{d}\right)^{n/2}\Ex_{U\sim\muU} \left[ P_U(x) P_U(y) \right].\label{eq:clif_via_haar}
\end{equation}
Note that this upper bound is also given in terms of the corresponding Haar value $\Ex_{\muU} \left[ P_U(x)P_U(y)\right]$. We use the second moment already calculated in \Cref{eq:Clifford_second_mom}. Finally, we bound the prefactor: 
By Bernoulli's inequality, we have that $(1+x^d)^n \leq e^{nx^d}$. For $d\geq \frac{\log n}{\log 5/4}$ and $x<1$ we can then use the convexity of the exponential function $e^{y}\leq(1-y)e^0+ye^1$ to obtain $e^{nx^d}\leq 1-nx^d + enx^d\leq 1+2nx^d$. This allows us to show that 
\begin{equation}
	\left(1+\left(\frac{4}{5}\right)^{d}\right)^{n/2} \leq 1 + n \left(\frac{4}{5}\right)^{d}.\label{eq:prefactor}
\end{equation}

\noindent Substituting \Cref{eq:Clifford_second_mom,eq:prefactor} into \Cref{eq:clif_via_haar} then yields \Cref{eq:brickwork_second_mom}.
\end{proof}

\section{Random Clifford unitaries}\label{app:far-from-uniform-global-cliffords}

The Clifford group forms a $3$-design. Therefore, we can carry over the bounds on $\f$ obtained via Chebychev and hence second moments from the global Haar measure to the uniform measure over global Clifford operations. 
The same analogy holds for local Haar random unitaries and local Clifford unitaries. 
Thus, the bounds on $\f$ from the restricted depth moments from \Cref{app:restricted-depth-moments} also hold for restricted depth Clifford circuits.
The key difference between the case of Clifford and Haar random unitaries lies thus in the far from uniform behavior. 
This is emphasized in the following lemma. 

\begin{lemma}
 The probability that the output distribution of a uniformly random global Clifford circuit on $n$ qubits is the uniform distribution is given by
\begin{equation}
    \Pr_{U\sim \mathrm{Cl}(2^n)} \left(P_U = \mathcal{U} \right) = \frac{1}{\prod_{i=1}^{n}\left(1+\frac{1}{2^{i}}\right)} \; .
\end{equation}
In particular, it asymptotically approaches
\begin{equation}
    \Pr_{U\sim \mathrm{Cl}(2^n)} \left(P_U = \mathcal{U} \right) \overset{n\to\infty}{\to} 0.41942244...
\end{equation}
from above and for any number of qubits $n$, the probability is larger than $0.41$.
\end{lemma}

Thus, even though ``non-trivial'' learning is hard for random Clifford unitary output distributions as characterized by $\f$, the trivial learning algorithm, which always returns $\mc U$, will succeed with probability larger than $0.41$ over the uniformly drawn $U\sim\Cl(2^n)$. 

\begin{proof}
The result of drawing a uniformly random Clifford unitary $U\sim \mathrm{Cl}(2^n)$ and applying to $\ket{0}^{\otimes n}$ is a uniformly random stabilizer state.

The number of $n$-qubit stabilizer states is given by \cite{aaronson_gottesman_2004}
\begin{equation}
    |\mathcal{S}_n| = 2^{n}\prod_{i=i}^{n}\left(2^{i}+1\right)= 2^{n+n(n+1)/2}\prod_{i=1}^{n}\left(1+\frac{1}{2^{i}}\right)
\end{equation}

The number of $n$-qubit stabilizer states giving rise to the uniform distribution is given by
\begin{equation}
    |\mathcal{S}^n_n| = 2^n\cdot 2{n(n+1)/2} = 2^{n+n(n+1)/2}
\end{equation}
This follows from \cite{kueng2015qubit}. In particular, Corollary 2 in Ref. \cite{kueng2015qubit} gives a formula for the number of stabilizer states with pre-described inner product with respect to a fixed reference stabilizer state. For our purposes, it suffices to take as reference state the all-zero state $\ket{0^n}$ and find the number of stabilizer states $\ket{\psi}$ such that $|\bra{\psi} \ket{0^n}|=2^{-n}$. Such states are precisely the $n$-qubit stabilizer states giving rise to the uniform distribution.

Hence, 
\begin{equation}
     \Pr_{U\sim \mathrm{Cl}(2^n)} \left(P_U = \mathcal{U} \right) = \frac{ 2^{n+n(n+1)/2}}{2^{n+n(n+1)/2}\prod_{i=1}^{n}\left(1+\frac{1}{2^{i}}\right)} = \frac{1}{\prod_{i=1}^{n}\left(1+\frac{1}{2^{i}}\right)}
\end{equation}

The asymptotic behavior of this product for $n\to \infty$ is found in
\cite{MSELimit}.
\end{proof}

\section{Deterministic algorithms}\label{app:omitted-proof-lemma}
The aim of this appendix is to give a detailed proof of \Cref{lem:deterministic-average-complexity-2} which is restated as \Cref{thm:deterministic-aqc-restated}. 
We follow a similar strategy as Feldman in \cite{feldman_general_2017} by proving the result for learning via a reduction to a suitably chosen decision problem.

\begin{problem}[Decide $\D$ versus $Q$]\label{prob:decide}
Let $\D$ be some distribution class and $Q$ some fixed reference distribution.
The task decide $\D$ versus $Q$ is defined as, given access to an unknown $P\in\D\cup\lrb{Q}$ to decide whether ``$P=Q$'' or ``$P\in\D$''.
\end{problem}

We connect the query complexity of learning with the query complexity of deciding by the following lemma. .

\begin{lemma}[Learning is as hard as deciding]\label{lem:learning-is-as-hard-as-deciding-deterministic}
Let $\D$ be a distribution class and let $Q$ be such that $\tv(P,Q)>\epsilon+\tau$ for all $P\in\D$. 
Let $0<\tau\leq\epsilon\leq1$.
Let $\A$ be a deterministic algorithm that $\epsilon$-learns $\D$ from $q$ many $\tau$ accurate statistical queries. 
Then there exists a deterministic algorithm that decides $\D$ versus $Q$ from $q+1$ many $\tau$ accurate statistical queries.
\end{lemma}

\begin{proof}
We run $\A$ on the unknown distribution $P\in\D\cup\lrb{Q}$ and obtain either
\begin{itemize}
    \item a representation of some $P'$ which is $\epsilon$ close to $P$ if $P\in\D$, or
    \item anything if $P=Q$.   
\end{itemize}
In case we do not receive a representation of any distribution return ``$P=Q$''.
Now, assume we receive a representation of some distribution $P'$. Using this representation compute whether $P'$ is $\epsilon$ close to any distribution in $\D$. 
While this step is computationally costly, it does not require any further queries to $\stat(P)$.
If there does not exists such a distribution in $\D$ which is $\epsilon$-close to $P'$, return ``$P=Q$''.

Now assume there exists an $H\in\D$ such that $\tv(P',H)<\epsilon$. To assure, that $\A$ is not biased towards returning distributions close to $\D$ if it fails, compute the set $S$ that maximizes the total variation distance between $Q$ and $P'$, $\abs{P'(S)-Q(S)}=\tv(P',Q)$. Denote by $\phi=\mb1_S$ the characteristic function on $S$ and query $v_\phi\leftarrow\stat_\tau(P)[\phi]$.
If $\abs{Q[\phi]-v_\phi}\leq\tau$ return ``$P=Q$'', else return ``$P\in\D$''.

We analyze the algorithm for each case separately. Common to both is that the algorithm makes, by definition, at most $q+1$ statistical queries. 

To begin with assume $P\in\D$. By the correctness of $\A$ we receive a representation of some $P'$ that is at most $\epsilon$ far from $P$. By assumption, for any $H\in\D$ it holds $\tv(H,Q)>\epsilon+\tau$.
Then, by the definition of $S$  using the reverse triangle inequality we find 
\begin{align}
    \abs{Q[\phi]-v_\phi}\geq
    \abs{\abs{Q[\phi]-P[\phi]}-\abs{P[\phi]-v_\phi}}>
    \abs{\epsilon+\tau-\tau}
    =\epsilon\geq\tau\,.
\end{align}
Hence, we correctly decide ``$P\in\D$''.

For the other case assume $P=Q$. If $\A$ does not return a valid representation or, if $\A$ returns a representation of some $P'$ that is more than $\epsilon$ far away from any distribution in $\D$, we know, by the correctness of $\A$, that it must hold $P=Q$. 
It remains to show the last step. Assume there is an $H\in\D$ which is at most $\epsilon$ far from $P'$. Then, by assumption, for every $\phi$ it must hold $\abs{Q[\phi]-v_\phi}\leq\tau$. Hence, we correctly decide ``$P=Q$''.
\end{proof}
% \MI{Is this lemma really different to our lemma 1 in our earlier paper? The archiv one.}
Using \Cref{lem:learning-is-as-hard-as-deciding-deterministic} it will be sufficient to bound the query complexity of deciding. This is achieved by the next lemma.

\begin{lemma}[Hardness of deciding, deterministic version]\label{lemma:hardness-of-deciding-deterministic}
Let $\A$ be a deterministic algorithm that decides $\D$ versus $Q$ from $q$ many $\tau$-accurate statistical queries, then for any measure $\mu$ over $\D$ it holds
\begin{align}
    q\geq \lr{\max_\phi\Pr_{P\sim\mu}\lrq{\abs{P[\phi]-Q[\phi]}>\tau}}^{-1}\,.
\end{align}
\end{lemma}

\begin{proof}

Assume we run $\A$ and answer every query $\phi$ by $Q[\phi]$. Denote by $\phi_1,\dots,\phi_q$ the queries made.
Assume for a contradiction, that for some $P\in\D$ there does not exist any distinguishing query. 
Then, the responses $Q[\phi_i]$ for $i=1,\dots,q$ would be valid responses for some $\stat_\tau(P)$ contradicting the assumption that $\A$ successfully decides whether $P=Q$.
Thus, for any $P\in\D$ there must exist at least one $i$ that distinguishes $Q$ from $P$.
In particular,
\begin{align}
    1 &= \Pr_{P\sim\mu}\lrq{\exists i,\,\abs{P[\phi_i]-Q[\phi_i]}>\tau}\label{eq:existence}\\
    &\leq \sum_{i=1}^q \Pr_{P\sim\mu}\lrq{\abs{P[\phi_i]-Q[\phi_i]}>\tau}\label{eq:after_union}\\
    &\leq q\max_\phi\Pr_{P\sim\mu}\lrq{\abs{P[\phi_i]-Q[\phi_i]}>\tau}\,,
\end{align}
which completes the proof.
\end{proof}

We are now set to prove our bound for the deterministic average case query complexity. Note, that \Cref{lem:learning-is-as-hard-as-deciding-deterministic} holds for learners that learn all of $\D$. Thus, the core of the remaining proof will be to translate the implications on worst to average case learners.

\begin{theorem}[\Cref{lem:deterministic-average-complexity-2} restated]\label{thm:deterministic-aqc-restated}
Suppose there is a deterministic algorithm $\A$ that $\epsilon$-learns a $\beta$ fraction of  $\D$ with respect to $\mu$ from $q$ many $\tau$-accurate statistical queries. Then for any $Q$ it holds
\begin{align}
    q+1\geq\frac{\beta-\Pr_{P\sim\mu}\lrq{\tv(P,Q)\leq\epsilon+\tau}}{\max_\phi\Pr_{P\sim\mu}\lrq{\abs{P[\phi]-Q[\phi]}>\tau}}\,,
\end{align}
where again, the $\max$ is over all functions $\phi:X\to[-1,1]$.
\end{theorem}

\begin{proof}
Let $\D'\subseteq\D$ with $\mu(\D')=\beta$ be a set on which $\A$ is successful.
Define 
\begin{align*}
    \D_Q=\lrb{P\in\D':\tv(P,Q)>\epsilon+\tau}
\end{align*}
and let $\mu_Q$ be the measure $\mu$ conditioned on $\D_Q$, such that $\mu_Q(P)=\mu(P\mid P\in\D_Q)$. 
Then, by the definition of the conditional probability,
\begin{align}
    \beta-\Pr_{P\sim\mu}\lrq{\tv(P,Q)\leq\epsilon+\tau}\leq\mu(\D_Q)\,.
\end{align}
Therefore, for any $\phi$ it holds
\begin{align}
\begin{split}
    \Pr_{P\sim\mu_Q}\lrq{\abs{P[\phi]-Q[\phi]}>\tau} = \Pr_{P\sim\mu}\lrq{\abs{P[\phi]-Q[\phi]}>\tau\mid P\in\D_Q} &\\
    \leq 
    \frac{\Pr_{P\sim\mu}\lrq{\abs{P[\phi]-Q[\phi]}>\tau}}{\beta-\Pr_{P\sim\mu}\lrq{\tv(P,Q)\leq\epsilon+\tau}}&
\end{split}
\end{align}
where we used the definition of the conditional probability.

The claim then follows from the observation that the average learner $\A$ for $\D$ implies a learner for $\D_Q$, the complexity of which can be bounded by the complexity of deciding $\D_Q$ vs $Q$ via the reduction \Cref{lem:learning-is-as-hard-as-deciding}. 
We obtain a bound for the latter from \Cref{lem:hardness-of-deciding} applying the measure $\mu_Q$.
\end{proof}

Before we end this section we state a variant of this bound due to Feldman to discuss the connection to \Cref{thm:deterministic-aqc-restated}. We restate his proof adapted to our notation for the sake of completeness.

\begin{lemma}[Variation of Lemma C.2 from \cite{feldman_general_2017} for deterministic algorithms]\label{lem:c2}
Suppose there is a deterministic algorithm $\A$ that $\epsilon$-learns a $\beta$ fraction of $\D$ with respect to $\mu$ from $q$ many $\tau$-accurate statistical queries. Then for any $Q$ it holds
\begin{align}
    q\geq\frac{\beta-\sup_{D}\Pr_{P\sim\mu}\lrq{\tv(P,D)<\epsilon}}{\max_\phi\Pr_{P\sim\mu}\lrq{\abs{P[\phi]-Q[\phi]}>\tau}}\,,
\end{align}
where the $\max$ is over all functions $\phi:X\to[-1,1]$ and the $\sup$ is over all distributions $D$ over the domain $X$.
\end{lemma}

\begin{proof}
Denote by $\D'\subseteq\D$ the subset of size $\mu(\D')=\beta$ on which $\A$ successfully $\epsilon$-learns from $q$ queries.  
We run $\A$ and answer every query $\phi$ by $Q[\phi]$.
By assumption $\A$ makes $q$ queries $\phi_1,\dots,\phi_q$ to $Q$ and, without loss of generality, we assume that the algorithm returns some distribution $D$.
Now, let $P$ be any distribution in $\D'$ at least $\epsilon$-far from $D$.
Exactly as in the proof of \Cref{lemma:hardness-of-deciding-deterministic}, there must exists at least one query function $\phi_i$ that distinguishes $Q$ from $P$.
In particular, it must hold
\begin{align}
\begin{split}
    \beta-\Pr_{P\sim\mu}\lrq{\tv(P,D)<\epsilon} 
    &\leq \Pr_{P\sim\mu} \lrq{\exists i :\,\abs{P[\phi_i]-Q[\phi_i]}>\tau}\\
    &\leq \sum_{i=1}^q \Pr_{P\sim\mu}\lrq{\abs{P{\phi_i]-Q[\phi_i]}>\tau}}\\
    &\leq q\max_\phi\Pr_{P\sim\mu}\lrq{\abs{P[\phi]-Q[\phi]}>\tau}\,.
\end{split}
\end{align}
Now assume that, after interacting with $Q$, the algorithm does not return any valid distribution. 
Then, again by $\A$'s determinism, for \textit{any} $P\in\D'$ there must exist a distinguishing query $\phi_i$ that distinguishes $Q$ from $P$.
The claim then follows by taking the supremum over $D\in\D_X$ to bound the $\epsilon$-ball around the unknown $D$.
\end{proof}

\begin{note}
We want to highlight that \Cref{lem:c2} is tight with respect to $\beta$: The trivial algorithm, which makes zero queries and always outputs that $D$ which maximizes the open $\epsilon$-ball will, with probability $\mu(B_\epsilon(D))$ over $P\sim\mu$ be correct.
\end{note}

\begin{note}\label{note:whythiswhythat}
As stated above, \Cref{lem:c2} gives the optimal lower bound with respect to $\beta$. 
However, in some cases directly bounding the weight of all $\epsilon$-balls may not be convenient.
In \Cref{app:far-from-everywhere} we give a general recipe for obtaining bounds for all $\epsilon$-balls just from the two ingredients used in \Cref{thm:deterministic-aqc-restated}: The maximally distinguishable fraction and the mass of the ball around the reference distribution.
While this strategy is straight forward, the bounds obtained are slightly worse than those obtained by directly invoking \Cref{thm:deterministic-aqc-restated}, which is why we stick to the latter result in this work. 
\end{note}

\section{Quantum and probabilistic algorithms}\label{app:randomized}

In this appendix we will detail the connection between statistical query learning via deterministic and random algorithms. 
Throughout the section we will refer by random algorithm to 
both classical probabilistic as well as quantum algorithms. 

The randomized average case query complexity for $\epsilon$-learning $\D$ with respect to the probability measure $\mu$ depends on the two parameters $\alpha$ and $\beta$, where
\begin{itemize}
    \item $\alpha$ denotes the success probability with respect to the internal randomness of the learning algorithm and
    \item $\beta$ denotes the fraction of distributions in $\D$ measured with respect to $\mu$ on which the learning algorithm must be successful. 
\end{itemize}

The aim of this appendix is to bound the randomized average case query complexity for $\epsilon$-learning $\D$ by (c.f. \Cref{thm:random-average-complexity})
\begin{align}
    q\geq \frac{2\cdot\lr{\alpha-\frac12}\cdot\lr{\beta-\Pr_{P\sim\mu}\lrq{\tv(P,Q)\leq\epsilon+\tau}}}{\max_\phi\Pr_{P\sim\mu}\lrq{\abs{P[\phi]-Q[\phi]}>\tau}}\,.
\end{align}
Thus, the randomized average case query complexity of $\epsilon$-learning is bounded by the same bounds as the deterministic average case query complexity up to a prefactor $2(\alpha-1/2)$, which becomes trivial for $\alpha\leq1/2$.

We will follow the same strategy as in the deterministic case laid out in \Cref{app:omitted-proof-lemma}. 
The main difference is that we need a bound on the decision problem \Cref{prob:decide} for random algorithms. 
The subtlety, why the arguments from \Cref{lemma:hardness-of-deciding-deterministic} fail, is that a random algorithm does not need to make a distinguishing query to solve the problem. 
Rather, it may guess the correct solution using its internal randomness.
The main technical ingredient of this Appendix is thus a result by Feldman which, first bounding the probability of guessing correctly, bounds the statistical query complexity for \Cref{prob:decide} which is stated as \Cref{lem:hardness-of-deciding}. 

To begin with, we can follow the same strategy as before to reduce deciding to learning, also in the random setting.

\begin{lemma}[Learning is as hard as deciding]\label{lem:learning-is-as-hard-as-deciding}
Let $\D$ be a distribution class and let $Q$ be such that $\tv(P,Q)>\epsilon+\tau$ for all $P\in\D$. 
Let $0<\tau\leq\epsilon\leq1$.
Let $\A$ be an algorithm that $\epsilon$-learns $\D$ from $q$ many $\tau$ accurate statistical queries with probability $\alpha$ over its internal randomness. 
Then there exists an algorithm that, with probability $\alpha$ over its internal randomness, decides $\D$ versus $Q$ from $q+1$ many $\tau$ accurate statistical queries.
\end{lemma}

\begin{proof}
    The proof is identical to that of \Cref{lem:learning-is-as-hard-as-deciding-deterministic}. 
    The only difference is that the learner $\A$ only succeeds with probability $\alpha$, which leads to the decider only succeeding with probability $\alpha$. 
    The reduction itself however is deterministic and, as such, does not change the statistics.
\end{proof}

We now state the result by Feldman on the randomized statistical query complexity of \Cref{prob:decide}. For the sake of self consistency we provide the proof adapted to our notation.

\begin{lemma}[Hardness of deciding. Taken from Theorem 3.9 from \cite{feldman_general_2017}]\label{lem:hardness-of-deciding}
Let $\A$ be a random algorithm that decides $\D$ versus $Q$ with probability at least $\alpha$ from $q$ many $\tau$ accurate statistical queries. Then, for any measure $\mu$ over $\D$ it holds
\begin{align}
    q\geq\frac{2\cdot(\alpha-\frac12)}{\max_\phi\Pr_{P\sim\mu}\lrq{\abs{P[\phi]-Q[\phi]}>\tau}}\,.
\end{align}
\end{lemma}

\begin{proof}
Let $\A$ be an algorithm that decides $\D$ vs. $Q$ with probability $\alpha$ over its internal randomness. 
We run $\A$ and, on every query $\phi$ return $Q[\phi]$ . Denote by $\phi_1,\dots,\phi_q$ the queries made. These queries then can be interpreted as random variables with respect to $\A$'s randomness.
Let $P\in\D$ and denote by
\begin{align}\label{eq:probability-of-distinguishing}
p(P)=\Pr_\A\lrq{\exists i\,:\,\abs{P[\phi_i]-Q[\phi_i]}>\tau}\,.
\end{align}
We now show that $p(P)\geq2(\alpha-1/2)$:
By the correctness of $\A$ the corresponding output will be ``$P\in\D$'' with probability at most $1-\alpha$.
For the sake of contradiction assume $p(P)<2(\alpha-1/2)$.
Thus, when run on $P$, for some valid answers $\A$ will still return ``$P=Q$'' with probability at least $>1-p(P)-(1-\alpha)>1-2\alpha+1-1+\alpha=1-\alpha$.
Since this probability is bounded by $1-\alpha$ we find a contradiction $\alpha<\alpha$.
Thus $p(P)\geq2(\alpha-1/2)$.

The remainder now follows from the union bound as in \Cref{lemma:hardness-of-deciding-deterministic}:
\begin{align}
\begin{split}
    2(\alpha-1/2)\leq p(P)&=\Pr_{\A,P\sim\mu}\lrq{\exists i\,:\,\abs{P[\phi_i]-Q[\phi_i]}>\tau}\\
    &\leq \sum_{i=1}^q \Pr_{\A,P\sim\mu}\lrq{\abs{P[\phi_i]-Q[\phi_i]}>\tau}\\
    &\leq q\max_\phi \Pr_{P\sim\mu}\lrq{\abs{P[\phi]-Q[\phi]}>\tau}\,. 
\end{split}
\end{align}
\end{proof}

Thus, following \Cref{app:omitted-proof-lemma}, we can state the main theorem of this appendix.

\begin{theorem}[Randomized average case query complexity of learning]\label{thm:random-average-complexity}
Let $\A$ be a random algorithm for average case $\epsilon$-learning $\D$ with respect to $\mu$ with parameters $\alpha$ and $\beta$ from $q$ many $\tau$ accurate statistical queries. Then for any $Q$ it holds
\begin{align}
    q+1\geq2\cdot\frac{(\alpha-\frac12)\cdot(\beta-\Pr_{P\sim\mu}\lrq{\tv(P,Q)\leq\epsilon+\tau})}{\max_\phi\Pr_{P\sim\mu}\lrq{\abs{P[\phi]-Q[\phi]}>\tau}}\,.
\end{align}
\end{theorem}

\begin{proof}
    The proof is identically to that of \Cref{thm:deterministic-aqc-restated} using \Cref{lem:learning-is-as-hard-as-deciding} and \Cref{lem:hardness-of-deciding} instead of \Cref{lem:learning-is-as-hard-as-deciding-deterministic} and \Cref{lemma:hardness-of-deciding-deterministic}.
\end{proof}

For the sake of context relating to the discussion in the end of \Cref{app:omitted-proof-lemma} and \cite[Lemma C.2]{feldman_general_2017}, we finish this appendix with two additional insights.

\begin{note}
If we restrict the result to probabilistic algorithms we can follow the argument from \cite[Lemma C.2]{feldman_general_2017}: One can make the randomness explicit by writing the random algorithm $\A$ as an ensemble of deterministic algorithms $\{\A_x\}$ with $x\sim\A$ the internal randomness. 
Then the randomized average case query complexity can be bounded by the deterministic average case query complexity replacing $\beta$ by $\alpha\cdot\beta$.  This yields
\begin{align}\label{eq:feldman-c2}
    q\geq \frac{\alpha\cdot\beta-\sup_{D}\Pr_{P\sim\mu}\lrq{\tv(P,D)<\epsilon}}{\max_\phi\Pr_{P\sim\mu}\lrq{\abs{P[\phi]-Q[\phi]}>\tau}}\,,
\end{align}
where, for the sake of transparency, we used Feldmans bound stated as \Cref{lem:c2} for the deterministic reference bound.
\end{note}

Note that \Cref{eq:feldman-c2} is tight with respect to $\alpha\cdot\beta$ and the joint measure $\mu\times\A$.
However, it has two disadvantages for our usecase.
First, it only holds for classical probabilistic algorithms, but not for other random algorithms such as quantum algorithms.
Second, we are interested in the average case hardness as in \Cref{def:averagecasecomplexity}. This means, we would like a statement that is tight with respect to $\beta$ with respect to $\mu$ only.
Thus, to obtain the corresponding tight bound for quantum algorithms, we add the following lemma.

\begin{lemma}\label{lem:beta-tight-randomized-average-complexity-bound}
Let $\A$ be a random algorithm for average case $\epsilon$-learning $\D$ with respect to $\mu$ with parameters $\alpha$ and $\beta$ from $q$ many $\tau$-accurate statistical queries. Then for any $Q$ it holds
\begin{align}
    q\geq2\cdot\frac{(\alpha-\frac12)\cdot(\beta-\sup_D\Pr_{P\sim\mu}\lrq{\tv(P,D)<\epsilon})}{\max_\phi\Pr_{P\sim\mu}\lrq{\abs{P[\phi]-Q[\phi]}>\tau}}\,,
\end{align}
where the $\sup$ is over all distributions with the same domain $X$ and the $\max$ is over all functions $\phi:X\to[-1,1]$.
\end{lemma}

\begin{proof}
Assume $\A$ is a random algorithm that $\epsilon$-learns $\D$ with respect to $\mu$, $\alpha$ and $\beta$ from $q$ many $\tau$-accurate statistical queries.
We run $\A$ and answer each query $\phi$ by $Q[\phi]$. 
Denote by $\phi_1,\dots,\phi_q$ the queries made and, without loss of generality, assume $\A$ returns the representation of some distribution $D$. 
Denote by $\D'\subseteq\D$ the set on which, with probability at least $\alpha$, the algorithm is successful.
Further, let $p(P)$ as in the proof of \Cref{lem:hardness-of-deciding} and let 
\begin{align*}
    \D_D=\lrb{P\in\D':\tv(P,D)\geq\epsilon}\,.
\end{align*}
Since $p(P)\geq2(\alpha-1/2)$ for any $P\in\D_D\subseteq\D'$ we find
\begin{align}
\begin{split}
    2&(\alpha-1/2)\leq\Pr_{P\sim\mu, \A}\lrq{\exists i\;,\;\abs{P[\phi_i]-Q[\phi_i]}>\tau\mid P\in\D_D} \\[10pt]
&= \frac{\Pr_{P\sim\mu, \A}\lrq{\exists i\;,\;\abs{P[\phi_i]-Q[\phi_i]}>\tau}}{\mu(\D_D)}\leq \frac{\Pr_{P\sim\mu, \A}\lrq{\exists i\;,\;\abs{P[\phi_i]-Q[\phi_i]}>\tau}}{\beta-\Pr_{P\sim\mu}\lrq{\tv(P,D)<\epsilon}}\\[10pt]
&\leq \sum_{i=1}^q\frac{\Pr_{P\sim\mu, \A}\lrq{\abs{P[\phi_i]-Q[\phi_i]}>\tau}}{\beta-\Pr_{P\sim\mu}\lrq{\tv(P,D)<\epsilon}}\leq q\frac{\max_\phi\Pr_{P\sim\mu}\lrq{\abs{P[\phi]-Q[\phi]}>\tau}}{\beta-\Pr_{P\sim\mu}\lrq{\tv(P,D)<\epsilon}}\,,
\end{split}
\end{align}
where we used the definition of the conditional probability, the bound on $\mu(\D_D)$ similar to that on $\mu(\D_Q)$ from the proof of \Cref{thm:deterministic-aqc-restated} and the union bound.
The claim then follows from taking the maximum over all distributions in order to estimate the unknown $D$.
\end{proof}

It is easy to see that \Cref{lem:beta-tight-randomized-average-complexity-bound} is tight with respect to $\beta$: The trivial algorithm that always returns $D$, where $D$ is the distribution with the $\epsilon$-ball of highest weight, will succeed with probability $\Pr_{P\sim\mu}\lrq{\tv(P,D)<\epsilon}$. 
We conclude this appendix with a note similar to \Cref{note:whythiswhythat}.

\begin{note}
In general \Cref{lem:beta-tight-randomized-average-complexity-bound} gives the optimal lower bounds with respect to $\beta$. 
However, in some cases directly bounding the weight of all $\epsilon$-balls may not be convenient.
The following appendix \Cref{app:far-from-everywhere} gives a general recipe for obtaining such a bound just from the two ingredients used in \Cref{thm:random-average-complexity}: The maximally distinguishable fraction and the mass of the ball around the reference distribution.
While this strategy is straight forward, the bounds obtained are slightly worse than those obtained by directly invoking \Cref{thm:random-average-complexity}, which is why we stick to the latter result in this work. 
\end{note}

\section{Far from any fixed distribution}
\label{app:far-from-everywhere}
In the main text, we obtained multiple ``far-from-uniform"-results for the output distributions of random circuits for different depth regimes. In this section, we show that random quantum circuits actually exhibit a more general property. 
Namely, their output distributions are with overwhelming probability far from any fixed distribution. This result was stated in the main text as \Cref{ithm:farfromeverything}. Here, we restate it formally and then go on to prove it.

\begin{theorem}[Formal version of \Cref{ithm:farfromeverything}]\label{thm:farfromeverything}
Let $\muC$ be the measure on $\U(2^n)$ induced by local random quantum circuits of depth $d$. Then, there is a $d'=O(n)$ such that at any depth $d\geq d'$, for any $\epsilon \leq   1/225$, and for any distribution $D$ over $\{0,1\}^n$ it holds
\begin{align}
    \Pr_{U\sim\muC}\lrq{\tv(P_U,D)\geq \epsilon}\geq 1-c2^{-n}
    \,,
\end{align}
where $c$ is a constant that can be bounded by $c<7\times10^6 < 2^{20}$.
\end{theorem}

In the following, let $D$ denote the arbitrary but fixed distribution as in the above theorem. We note that to prove \Cref{thm:farfromeverything}, we can distinguish two cases: Either $D$ is itself close to uniform, then a far-from-uniform bound implies a far-from-$D$ bound. 
In the other case, $D$ is at least some distance away from the uniform distribution.
 As made explicit by the following lemma, the ``far-from-any-fixed-distribution"-result for such $D$ is implied by a bound on the maximal distinguishable fraction with respect to the uniform distribution, $\f=\kfrac(\mu,\mc U,\tau)$. In fact, the lemma holds not only for the uniform distribution but any choice of reference distribution $Q$.
 
 Thus, a ``far-from-any-fixed-distribution"-result follows from two ingredients: A ``far-from-$Q$" result and a bound on the maximally distinguishable fraction against $Q$, for any reference distribution $Q$. We  happen to have calculated these bounds for the particular choice of $Q=\mc U$ already in the main text.

\begin{lemma}\label{lem:farfromeverything_1}
Let $\epsilon,\tau>0$, $X$ be some domain and let $Q\in\D_X$ be the \textit{reference} distribution. Moreover, let $D\in\D_X$ be such that
\begin{equation}
\tv\left(Q,D \right)>\epsilon+\tau\,.
\end{equation}
Then for any measure $\mu$ over $\D_X$ it holds
\begin{align}
    \Pr_{P\sim\mu}\lrq{\tv(P,D)<\epsilon}\leq\kfrac(\mu, Q, \tau)\,.
\end{align}
\end{lemma}

\begin{proof}
Recall that by the variational characterization of the total variation distance it holds that $\tv(Q,D)=\max_{T\subset X}\abs{Q(T)-D(T)}$.
Let $S\subseteq X$ be such a set maximizing the total variation distance 
and denote by $\phi=\mb1_S$ the characteristic function on $S$.
This is, $\tv(D,Q)=\abs{D[\phi]-Q[\phi]}$.

By the reverse triangle inequality for any $P'\in B_\epsilon(D)$ it then holds
\begin{align}
\begin{split}\label{eq:balltotauestimation}
    \abs{P'[\phi]-Q[\phi]}&\geq 
    \abs{\abs{P'[\phi]-D[\phi]} - \abs{D[\phi]-Q[\phi]}}\\
    &\geq\abs{\tv(Q,D) - \tv(P',D)}\\
    &>\abs{\epsilon+\tau-\epsilon}=\tau\,,
\end{split}
\end{align}
where we have used that $\abs{P'[\phi]-D[\phi]}\leq\tv(P',D)<\epsilon$ together with $\tv(Q,D)>\epsilon+\tau>\epsilon>\tv(P',D)$.

Hence, by \Cref{eq:balltotauestimation} it holds
\begin{align}
\Pr_{P\sim\mu}\lrq{\tv(P,D)<\epsilon}\leq \Pr_{P\sim\mu}\lrq{\abs{P[\phi]-Q[\phi]}>\tau}\leq\kfrac(\mu,Q,\tau)\,,
\end{align}
where the last inequality is due to the maximum over all functions $\phi$ in the definition of $\kfrac(\mu,Q,\tau)$ (\Cref{def:frac}). 
\end{proof}

Applying \Cref{lem:farfromeverything_1} to the choice of $Q=\mc U$ and using our bound for the maximally distinguishable fraction $\f$ against uniform from \Cref{l:frac_linear_depth} in \Cref{sec:frac-linear-depth}, we find the following:

\begin{corollary}
\label{cor:farfromeverything}
Let $\epsilon,\tau>0$ and let $D$ be any probability distribution
over $\left\{ 0,1\right\} ^{n}$ satisfying
\begin{equation}
\tv\left(D,\mathcal{U}\right)>\epsilon+\tau
\end{equation}
where $\mathcal{U}$ is the uniform distribution. Let $\muC$ denote the measure over brickwork random quantum circuits as in \Cref{def:rqc}.
Then, there is a $d'=O(n)$ such that at any depth $d\geq d'$ it holds
\begin{equation}
\Pr_{U\sim\muC}\left[\tv(D, P_U) < \epsilon \right]
\leq \frac{3}{2^{n}\tau^{2}}.
\end{equation}
\end{corollary}

We now complete the proof of \Cref{thm:farfromeverything} following the two-cases argument laid out above.

\begin{proof}[Proof of \Cref{thm:farfromeverything}]
Let $D$ be an arbitrary distribution, let $d$ large enough and let $\epsilon = 1/450$ such that $3\epsilon=\epsilon'=1/150$. We distinguish two cases:
\begin{enumerate}
    \item $\tv(D,\mathcal{U}) \leq 2\epsilon$,
    \item $\tv(D,\mathcal{U})  > 2\epsilon$.
\end{enumerate}

In case 1, we have that
\begin{equation}
    \Pr_{U\sim \muC}\left[\tv(P_{U},D) < \epsilon \right]
    \leq \Pr_{U\sim \muC} \left[\tv \left(P_{U},\mc U \right) < \epsilon' \right]
    <3200\cdot 2^{-n}
    \leq
    O\left( 2^{-n}\right)
\end{equation}
by our far-from-uniform result from the main text, namely \Cref{cor:far-from-uniform-linear}.

In case 2,  we have that $\tv(D,\mathcal{U})>2\epsilon$. Setting $\tau=\epsilon$ we can apply \Cref{cor:farfromeverything} 
to obtain 
\begin{equation}
    \Pr_{U\sim \muC}\left[\tv(P_{U},D) < \epsilon \right]
    <
    \frac{3}{2^n\epsilon^2}=607500\times 2^{-n}
    \,.
\end{equation}

Hence,   for any distribution $D$, any $\epsilon  \leq 1/450$ we find
\begin{equation}
     \Pr_{U\sim \muC}\left[\tv(P_{U},D) < \epsilon \right]
     < 607500\times 2^{-n}
     =O\left(2^{-n} \right)\,.
\end{equation}
\end{proof}

Finally, we summarize the connection between the $\epsilon$-ball with the largest weight and the maximally distinguishable fraction as advertised at the end of \Cref{app:omitted-proof-lemma} as follows.

\begin{lemma}\label{lem:farfromeverything_2}
Let $\epsilon,\tau>0$, $X$ be some domain, $\mu$ be a measure over $\D_X$ and $Q\in\D_X$ an arbitrary distribution. Then for any $D\in\D_X$ it holds
\begin{align}
\begin{split}
    \Pr_{P\sim\mu}\lrq{\tv(P,D)<\epsilon}\leq 
    \max\lrb{\kfrac(\mu,Q,\tau)\,,\,\, \Pr_{P\sim\mu}\lrq{\tv(P,Q)\leq2\epsilon+\tau} }\,.
\end{split}
\end{align}
\end{lemma}

\begin{proof}
We consider two cases. In case $\tv(D,Q)>\epsilon+\tau$ 
we obtain the contribution from $\kfrac(\mu,Q,\tau)$  
via \Cref{lem:farfromeverything_1}. So consider the case $\tv(D,Q)\leq \epsilon+\tau$. In this case we can bound 
\begin{align}
\Pr_{P\sim\mu}\lrq{\tv(P,D)<\epsilon}\leq\Pr_{P\sim\mu}\lrq{\tv(P,Q)\leq2\epsilon+\tau}\,,
\end{align}
which yields the claim.
\end{proof}

%%%%%%
%
%  BIBLIOGRAPHY
%
%%%%%%

\Urlmuskip=0mu plus 1mu\relax
\printbibliography

\end{document}